\documentclass[review]{elsarticle}[12pt]

\usepackage{lineno,hyperref}
\usepackage{graphicx}
\usepackage{caption}
\usepackage{subcaption}
\makeatletter
\def\ps@pprintTitle{%
  \let\@oddhead\@empty
  \let\@evenhead\@empty
  \def\@oddfoot{\reset@font\hfil\thepage\hfil}
  \let\@evenfoot\@oddfoot
}
\makeatother

\begin{document}

\begin{frontmatter}

\title{Limited Fetch Revisited: Comparison of Wind Input Terms, in Surface Wave Modeling}

\author[a2,a3,a4]{Pushkarev Andrei \fnref{fn1} }

\author[a1,a2,a3,a4]{Zakharov Vladimir}

\address[a1]{University of Arizona, 617 N. Santa Rita Ave., Tucson, AZ 85721, USA}
\address[a2]{LPI, Leninsky Pr. 53, Moscow, 119991 Russia}
\address[a3]{Pirogova 2, Novosibirsk State University, Novosibirsk, 630090 Russia}
\address[a4]{Waves and Solitons LLC, 1719 W. Marlette Ave., Phoenix, AZ 85015 USA}

\fntext[fn1]{Corresponding author, \textit{E-mail}: dr.push@gmail.com}

\begin{abstract}

Results pertaining to numerical solutions of the Hasselmann kinetic equation ($HE$), for wind driven sea spectra, in the fetch limited geometry, are presented.
Five versions of source functions, including the recently introduced $ZRP$ model \cite{R9}, have been studied, for the exact expression of $S_{nl}$ and high-frequency implicit dissipation, due to wave-breaking. Four of the five experiments were done in the absence of spectral peak dissipation for various $S_{in}$ terms. They demonstrated the dominance of quadruplet wave-wave interaction, in the energy balance, and the formation of self-similar regimes, of unlimited wave energy growth, along the fetch. Between them was the $ZRP$ model, which strongly agreed with dozens of field observations performed in the seas and lakes, since 1947. The fifth, the $WAM3$ wind input term experiment, used additional spectral peak dissipation and reproduced the results of a previous, similar, numerical simulation described in \cite{R11}, but only supported the field experiments for moderate fetches, demonstrating a total energy saturation at half of that of the Pierson-Moscowits limit. The alternative framework for $HE$ numerical simulation is proposed, along with a set of tests, allowing one to select physically-justified source terms. 

\end{abstract}

\begin{keyword}
\texttt{Hasselmann equation, wind-wave interaction, wave-breaking dissipation, nonlinear interaction, self-similar solutions, Kolmogorov-Zakharov spectra}
\end{keyword}

\end{frontmatter}

%\linenumbers

\section{Introduction}

\noindent The motivation, for the research presented in the current paper, was to continue the project of finding a firm scientific foundation for the study of wind driven seas.

\noindent The most important step in this direction was made in 1962, by K.Hasselmann \cite{R65,R66,R113}, who proposed the kinetic equation for wind waves description

\begin{equation}
\label{HE}
\frac{\partial \varepsilon }{\partial t} +\frac{\partial \omega _{k} }{\partial \vec{k}} \frac{\partial \varepsilon }{\partial \vec{r}} =S_{nl} +S_{in} +S_{diss}
\end{equation}
similar to equations used in condensed media physics since the 1920's \cite{R76}, where $\varepsilon =\varepsilon (\omega_k,\theta,\vec{r},t)$ is the wave energy spectrum, as a function of wave dispersion $\omega_k =\omega (k)$, angle $\theta$, two-dimensional real space coordinate $\vec{r}=(x,y)$ and time $t$. $S_{nl}$, $S{}_{in}$ and $S_{diss}$ are nonlinear, wind input and wave-breaking dissipation terms, respectively. Hereafter, only the deep water case, $\omega=\sqrt{g k }$ is considered, where $g$ is the gravity acceleration and $k=|\vec{k}|$ is the absolute value of wavenumber $\vec{k}=(k_x,k_y)$. 

\noindent Eq.(\ref{HE}) is widely accepted in the oceanographic community \cite{R10,R11} and has several names. It is called the Boltzmann equation \cite{R11} (while this is not exactly correct), the energy balance equation \cite{R10}, and the radiation balance equation. We will call it the Hasselmann equation (hereafter $HE$) as a tribute to Hasselmann's pioneering work. At the least, this is consistent with part of the community \cite{R67}.

\noindent The right side of Eq.(\ref{HE}) consists of three terms. The $S_{nl}$ term is completely known. It was consistently derived from Euler equations and describes quadruplets of waves satisfying resonant conditions

\noindent
\begin{equation}
\begin{array}{r@{}l}
\vec{k} + \vec{k_1} &{}= \vec{k_2} + \vec{k_3} \\
\omega_{k}+\omega_{k_1} &{}= \omega_{k_2}+\omega_{k_3}
\end{array}
\end{equation}
\noindent In the papers \cite{R1,R2} we introduced the following splitting of the $S_{nl}$
\begin{equation}
\label{SnlSplit}
S_{nl}(\omega,\theta) = F(\omega,\theta) -\Gamma(\omega,\theta)\varepsilon(\omega,\theta)
\end{equation}
\noindent The explicit expressions for $F$ and $\Gamma$ are presented in the Appendix. The motivation for this splitting is very simple. The term $F(k)$, for any spectral distribution $\varepsilon(\omega,\theta)$, \underline{is an essentially positive function}. We will soon show that this fact is of fundamental importance.

\noindent Kinetic equations similar to the Hasselmann equation are routinely used in different areas of theoretical physics. In all cases, the first and central issue is the description of solutions to the stationary equation 

\noindent
\begin{equation}
\label{SnlZero}
S_{nl}(\omega, \theta) = 0
\end{equation}

\noindent Any solution of this equation can be presented in the form

\begin{equation}
\varepsilon(\omega, \theta) = \frac{F(\omega, \theta)}{\Gamma(\omega,\theta)}
\end{equation}

\noindent
As far as $\varepsilon(\omega,\theta) > 0$, for all solutions 

\begin{equation}
\Gamma(\omega, \theta) > 0
\end{equation}

\noindent The function $\Gamma$ also has another physical sense. In the presence of nonlinear wave ensemble, the dispersion law is undergoing the re-normalization

\noindent
\begin{equation}
\omega(k) \rightarrow \omega(k) + \Delta \omega(k)
\end{equation} 

\noindent The re-normalization has real and imaginary parts. The imaginary part is
\begin{equation}
\label{ImaginaryPart}
Im \Delta(\omega) = \frac{1}{2} \Gamma(\omega,\theta)
\end{equation}

\noindent Everybody knows that Eq.(\ref{SnlZero}) has solutions with thermodynamic equilibrium. 

\noindent There is Maxwell distribution in the kinetic gas theory, and Plank distribution in quantum statistical dynamics. Physicists believed, for a long time, that the thermodynamic equilibrium spectra are unique solutions of Eq.(\ref{SnlZero}). This is certainly true, if the entropy of a solution is finite. However,  Eq.(\ref{SnlZero}) has a broad class of solutions with infinite entropy, governed by fluxes of conservative quantities -- energy, momentum and wave action.

\noindent These solutions are now called $KZ$ (Kolmogorov-Zakharov) solutions and widely used in different areas of physics (see, for instance \cite{R109,R110,R111,R112}). The general theory of $KZ$ solutions is described in the monograph \cite{R43}.

\noindent A more advanced development is contained in the paper \cite{R1}. The discovery of $KZ$ spectra was recognized by the physical community, by awarding a Dirac medal in 2003, for this development. 

\noindent The first KZ solution was found by Zakharov and Filonenko in 1966 (the English version of  \cite{R41} was published in 1967). It is the isotropic solution of the stationary Hasselmann Eq.(\ref{SnlZero}) (the details are presented in section 4):

\begin{equation}
\label{KZenergySolution}
\varepsilon(\omega) = \frac{\beta_{KZ}}{\omega^4} = C_K \frac{g^{4/3} P^{1/3}}{\omega^4}
\end{equation}

\noindent 
Here $P$ is the energy flux to the high frequency region. It was soon established, \cite{R107}, that the solution  \ref{KZenergySolution} is only ``the tip of the iceberg".  Actually,   Eq.(\ref{SnlZero}) has a much bigger class of $KZ$ solutions, outlined in the paper \cite{R1,R107}. The most interesting and important solutions, governed by fluxes of energy and momentum, are anisotropic. They are not exactly power-like, seeing their $\omega$ - dependence deviates from the $\omega^{-4}$ law, but only mildly. 

\noindent Meanwhile, numerous laboratory and field experiments showed that, in the important band of frequency, right behind the spectral peak (approximately for $1.5 \omega_p  <  3.5 \omega_p $), the observed spectra are very close to the $\omega^{-4}$ law. Experimental data obtained before 1985 was summarized in the well known paper of Phillips \cite{R59}. Since then, a lot of new data has accumulated (see, for instance, \cite{R101}, \cite{R47,R48,R50}, \cite{R7,R8}). Some other experimental results were cited in the article \cite{R108}.

 \noindent Recall that the exact $S_{nl}$ can be derived, rigorously, from the Euler equation.

\noindent Opposite to it, the ``source function" $S_{in}$ -- the energy income from the wind, and the energy dissipation function $S_{diss}$, due to wave-breaking, are only known approximately. In the oceanographic community, there is no consensus regarding their form. We discuss these questions in sections \ref{CurrentState} and \ref{TwoScenarios} of this paper.

The ambiguity of their proper definitions presents the first major issue for wind wave theory, and hinders development of accurate operational models, as well.

\noindent The other important issue is connected with $S_{nl}$ collision term numerical simulation. It is the complex, non-linear, operator, with deep internal symmetries. Several $S_{nl}$ simulation algorithms are available, at the moment, for example: Webb-Resio-Tracy ($WRT$) \cite{R68,R69} (also, see important paper \cite{R115}), Lavrenov \cite{R74} and Masuda \cite{R75}. The Van-Vledder version of the $WRT$ algorithm \cite{R79} has already been included in the $Wavewatch III$ and $SWAN$ models, for more than a decade. 

\noindent All of the above algorithms provide reliable results, but are too slow to provide simultaneous $HE$ solutions of the Eq.(\ref{HE}) for tens of thousands of spatial points, faster than real time, as is required by operational wave forecasting. Because of this, existing operational models use much faster substitutes for $S_{nl}$, in the form of $DIA$ and its analogs. This is not fatal, as long as the number of quadruplet configurations used in $DIA$ is large enough. However, what is wrong is the commonly practiced ``tuning" of the $DIA$ algorithm parameters, in the operational models.

\noindent We must stress, however, that we do not discuss the good and bad sides of different modifications of $DIA$ models. The only results discussed are those obtained from the numerical algorithm for solving the exact Hasselmann equation. This code is a modification of the WRT algorithm . We hereby call it $XNL$.

\noindent We insist that a correct definition of the source function is necessary, and we assert that it is possible to perform these corrections, without new theoretical constructions or new difficult experiments. It is sufficient to use existing experimental data, in a proper way. For 68 years, starting from a well-known work of Sverdrup and Munk \cite{R53}, oceanographers have accumulated a plethora of experimental facts regarding wave growth rate, with respect to winds. Some of those facts were obtained in water tanks, but the most interesting facts come from ocean measurements.

\noindent Nowadays, the results of numerous measurements for ``fetch limited" field set-ups, where the off shore wind and the waves are quasi-stationary, have been systematized and published \cite{R6}. 

\noindent All of those situations are described by the stationary $HE$ 

\noindent 
\begin{equation}
\label{HEstat}
\frac{\partial\omega}{\partial k}\frac{\partial \varepsilon}{\partial x} = S_{nl} + S_{in} + S_{diss}
\end{equation}

\noindent This equation is solved, in the presented research, for different source functions $S_{in}$ and $S_{diss}$. Five experiments were carried out, for different wind input functions, and their results were compared to known ocean field experimental data. This comparison actively used the fact that the results of those experiments are well described by {\it Weak Turbulence Theory} ($WTT$). This theory is explained, in detail, in the monograph \cite{R43}, and applications of this theory, to ocean experiments, are presented in the publications \cite{R4,R3,R5,R6,R7,R8}.

\noindent The possibility of $WTT$ application is based on the fact that, in Eq.(\ref{HE}), $S_{nl}$ is the dominant term. This fact can be explained in the following way. All $S_{in}$ cases considered in the current research are quasi-linear, which means that

\noindent
\begin{eqnarray}
S_{in} &=& \gamma (\omega, \theta)\varepsilon(\omega, \theta) \\
S_{diss} &=& -\gamma_{diss} (\omega,\theta) \varepsilon( \omega, \theta)
\end{eqnarray} 

\noindent Taking into account $S_{nl}$ splitting, Eqs.(\ref{SnlSplit}), (\ref{HEstat}) take the form

\noindent
\begin{equation}
\label{HEsplit}
\frac{\partial \omega}{\partial \vec{k}}\frac{\partial \varepsilon}{\partial \vec{r}} = F(\omega,\theta) - \left( \Gamma(\omega,\theta) - \gamma(\omega,\theta) + \gamma_{diss}(\omega,\theta) \right) \varepsilon(\omega,\theta) 
\end{equation}

\noindent One should note that $\gamma$ typically has a fairly small value of $10^{-5} \omega_p$, for waves with the frequencies close to the peak frequency $\omega_p$. The value of $\gamma_{diss}$ does not exceed $\gamma$, or waves are not excited at all. Meanwhile, the value of $\Gamma$ is rather large, as shown by analytic and numeric calculations. It easily exceeds $\gamma$,  by orders of magnitude \cite{R1,R2} (see Appendix). Therefore, one can substitute in the first approximation  Eq.(\ref{HEsplit}) by conservative equation 

\noindent
\begin{equation}
\label{HE_Cons}
\frac{\partial \omega_k}{\partial k}\frac{\partial \varepsilon}{\partial x} = S_{nl} 
\end{equation}

\noindent which is, indeed, the subject of the $WTT$ study.

\noindent It is customary to use ``Kitaigorodsky dimensionalization", where the fetch variable $x$, total energy $E$, and peak frequency $\omega_p$ are substituted by dimensionless variables
 
\noindent
\begin{equation}
\chi = \frac{xg}{u^2}, \,\,\, \varepsilon = \frac{E g^2}{u^2}, \,\,\, \hat \omega = \frac{\omega_p u}{g}
\end{equation}

\noindent All ocean and wave tank measurements show that $\varepsilon(\chi)$ and $\omega(\chi)$ are the power functions of dimensionless fetch $\chi$:

\noindent
\begin{eqnarray}
\label{Peq}
\varepsilon &= \varepsilon_0 \chi^p \\
\label{Qeq}
\hat \omega &= \omega_0 \chi^{-q}
\end{eqnarray}

\noindent The values of $p$ and $q$ vary in different experiments, but not significantly $0.74<p<1$, $0.2<q<0.3$. They are connected, with strong accuracy, by the ``magic relation" 

\noindent
\begin{equation}
\label{MagicRelation}
10q-2p=1
\end{equation}

\noindent These facts are explained by $WTT$ \cite{R3}. Conservative kinetic Eq.(\ref{HE_Cons}) has a 4-parameter family of self-similar solutions \cite{R3,R4,R5,R6}, for which the ``magic relation" is fulfilled exactly. 

\noindent It was shown in \cite{R9,R1} that the non-conservative $HE$, with the dissipation, localized in short waves, and forcing chosen as the power function 

\noindent
\begin{equation}
\gamma(\omega,\theta) = f(\theta)\omega^s
\end{equation}

\noindent also allows a self-similar solution and preserves the ``magic relation" Eq.(\ref{MagicRelation}).

\noindent All numerical experiments presented in the current paper included short-wave dissipation, but in the ``implicit" way: the spectrum at frequencies $f > 1.1 \,\,Hz$ was forced to Phillips spectrum $\varepsilon_\omega \simeq \omega^{-5}$. The validity of this approach is discussed in section \ref{TwoScenarios}. Similar procedure of matching the spectrum with the powerlike tail at high frequency is routine in the operational models \cite{R52}.

 \noindent Four out of five of those experiments assumed absence of long-wave dissipation. It is assumed hereafter that ``long waves" denotes the waves with the characteristic wavenumber close to the spectral peak vicinity.  Such experimental set-up contradicts existing tradition, but is justified by obtained results. Four existing wind forcing terms have been checked: $ZRP$ \cite{R9}, Chalikov \cite{R15,R25}, Hsiao-Shemdin \cite{R31} and Snyder \cite{R30}. The only  $S_{in}$ term, in the power form, was the $ZRP$ forcing term, and it was only this experiment which showed agreement with the field observation, for which $p=1$ and $q=0.3$. The other $S_{in}$ terms lead to the Eqs.(\ref{Peq}), (\ref{Qeq}), where indices $p$ and $q$ are functions of the dimensionless fetch. It is important to note that the ``magic relation" Eq.(\ref{MagicRelation}) still holds, as well, which means that corresponding spectra exhibit ``local" self-similarity.

\noindent The question is: how valid is the claim that the wave energy and the mean frequency behave like powers?  

\noindent The developed $WTT$ (Weak Turbulent Theory) does not impose limitations on wave energy growth, with the fetch. The maximum length of the dimensionless fetch $\chi\simeq 10^5$ was considered in the experiments of Donelan at al. \cite{R39}. The corresponding wave energy maximum was $\varepsilon = 4.07 \cdot 10^{-3}$, without any deviation from the power law. 

\noindent We should stress that we now speak about a stationary-in-time wave field, and discuss dependence of its characteristics on the fetch, only. In 1964, Person and Moscowitz formulated the hypothesis that on long enough fetches, the wave field becomes statistically homogeneous. According to Young \cite{R10} this ``spatial saturation" occurs at $\chi \simeq 5 \cdot 10^4 $ (see Fig.5.10 in the cited book) and the energy spectrum is stabilized on the level $\varepsilon \simeq 3.64 \cdot 10^{-3}$. For that time, it was an important achievement, but we must emphasize that this hypothesis was pure speculative, because Pierson and Moscowitz measured the wave field only at one spatial point. However, recent analysis of numerous experimental data, published in \cite{R6,R4,R5}, does not support the concept of the spatial saturation. Nevertheless, the idea of a ``mature sea" was actively supported in the paper \cite{R54} and became a kind of credo for oceanographers. The WAM3 model was designed specifically to support this idea.

\noindent The numerical test of the $WAM3$ model, using the $XNL$ approach, was performed in the fetch-limited geometry, long ago (see \cite{R11}, pp.229, 254, Fig.3.7 and Fig.3.22). It was found that for  moderate fetches $10^2<\chi<10^3$ this model describes the experimental situation pretty well, however, predicting saturation at the fetch $\chi \simeq 2\cdot 10^{4}$ on the low level $\varepsilon_{Max}\simeq 1.8 \cdot 10^{-3}$. Our numerical experiments confirmed these results. Moreover, we found that in the "practical fetch" interval $10^2 < \chi < 10^3$, the results of $WAM3$ coincide with the results obtained via the $ZRP$ model, without any spectral peak dissipation. For larger fetches, the $ZRP$ model demonstrates much better coincidence with field experiments, than the $WAM3$ model does.

\noindent Here one can recall William Okham's principle ``It is futile to do with more things that which can be done with fewer". Application of this principle leads to excluding the long wave dissipation from consideration, and to the conclusion that the $WAM3$ model is not consistent enough. It is satisfactory in only one aspect - it passes the $\omega^{-4}$ test, explained in section \ref{NumSetUp}.

\noindent The obtained results can be seen as a progressive step towards universal, physically-based, ocean surface wave models, the development of which will require minimal tuning for different ocean conditions. Other perspectives are discussed in the Conclusion.

\section{Current state of wind input source terms} \label{CurrentState}

\noindent Nowadays, the number of existing models of $S_{in}$ is large, but these models lack firm, theoretical, justification. Different theoretical approaches argue with each other. A detailed description of this discussion can be found in the monographs \cite{R10,R11}, and the papers \cite{R12,R13,R14,R15,R16,R31,R55,R39}.

\noindent The development of wind wave models had begun as far back as the 1920's, in the well-known works of Jeffreys \cite{R17}, \cite{R18}. His model is semi-empiric and includes an unknown ``sheltering coefficient". All other existing theoretical models are also semi-empiric, with one exclusion -- the famous Miles model \cite{R16}. This model is rigorous, but is related to an idealized situation -- the initial stage of wave excitation, by laminar wind, with specific wind profile $U(z)$.

\noindent Miles theory application is hampered by two circumstances. First is the fact that the atmospheric boundary layer is the turbulent one, and creating a rigorous, analytic theory of such turbulence is, as of today, an unsolvable problem.

\noindent There is the opinion, however, that wind speed turbulent pulsations are small, with respect to horizontal velocity $U(z)$, \cite{R20,R21,R22,R23} and that they should be neglected in the first approximation \cite{R21,R23}. This does not mean that turbulence is not taken into account, at all. It is suggesting that the role of the turbulence consists in formation of the averaged horizontal velocity profile. 

\noindent The wide spread opinion is that the horizontal velocity profile is distributed by the logarithmic law
\noindent
\begin{equation}
\label{LogProfile}
U(z) = 2.5 u_* \ln{}\frac{z}{z_0}
\end{equation} 
\noindent Here $u_*$ is the friction velocity and $z_0$ is the roughness parameter
\noindent
\begin{equation}
\label{Charnok}
z_0 = C_{ch} \frac{u_*^2}{g}
\end{equation} 
\noindent where $C_{ch} \simeq 3 \cdot 10^{-2}$ is the experimental and dimensionless Charnock constant.

\noindent One should note that the appearance of an anomalously small constant, not having ``formal justification", is an extremely rare phenomenon in physics. Eqs.(\ref{LogProfile}), (\ref{Charnok}) mean that the roughness parameter is very small: for typical ocean conditions -- wind speed $10$ m/sec on the height $z=10 \,\, m$ we get $z_0 \simeq 5 \cdot 10^{-4} \,\, m$. Such roughness is only twice the size of the viscid layer, defined from multiple experiments on turbulent wind flow, over smooth metal plates. Notice that the logarithmic law certainly could not work for a height of the order of few centimeters, where capillary effects are essential.

\noindent Usage of Eqs.(\ref{LogProfile}), (\ref{Charnok}) assumes, therefore, that the ocean behaves as smooth metal surface. This is not correct. Horizontal momentum is transferred to the smooth plate, on the surface itself, while in the ocean this process happens differently.

\noindent Momentum off-take, from the atmospheric boundary layer, is smoothly distributed over the whole width of the boundary layer and begins from the highest ``concurrence layer", i.e. from the height where the phase speed of the fastest wave matches the horizontal velocity.

\noindent Momentum off-take leads to horizontal velocity distribution $U(z)$ depending on time, the wave's development level, and energy spectrum. Meanwhile, Miles's instability increment is extremely sensitive to the horizontal velocity profile (there is no wave excitation for the linear profile $U(z)$, in Miles theory, for example). The velocity profile is especially important for slight elevations, on the order of several centimeters, over the water surface, which is almost unknown and difficult for experimental measurements. However, there have been some advances in this direction \cite{R19,R20}.

\noindent The necessity of taking into account the waves feedback, into the horizontal velocity profile, was understood a long time ago, as seen in the works of Fabrikant \cite{R21} and Nikolaeva at al. \cite{R22}. This approach was later continued by Jannsen \cite{R23} and explained, in detail, in the monograph \cite{R11} in the form of ``quasi-laminar" theory. This theory is lacking.

\noindent To consider the theory as self-consistent, even in the approximation of turbulence absence, it is necessary to solve equations describing the horizontal velocity profile $U(z)$, together with the Hasselmann equation, describing the energy spectrum evolution. This is not done yet, either.

\noindent That fact aside, many theoreticians do not share the opinion about turbulent pulsations insignificance, and consider them as the leading factor. Corresponding $TBH$ theory by Townsend, Belcher and Hunt \cite{R12} is an alternative to quasi-laminar theory. Both theories are discussed in \cite{R24}.

\noindent There is another approach, not connected with experimental analysis - numerical simulation of the boundary atmospheric layer, in the frame of empiric theories of turbulence. It was developed in the works \cite{R13,R14,R15,R25}. Since those theories are insufficiently substantiated, the same relates to the correspondingly derived wind input terms.

\noindent Across the wide variety of theoretical approaches to defining $S_{in}$, almost all of them are ``quasi-linear" \cite{R2} where the standard relation \cite{R10,R11}

\begin{eqnarray}
\label{GammaDefinition}
\gamma(\omega, \phi) = \frac{\rho_a}{\rho_w} \omega \beta(\frac{\omega}{\omega_0}, \phi)
\end{eqnarray} 

\noindent is being used. Here $\omega_0 = \frac{g}{u}$, where $u$ is the wind speed, defined differently in individual models. The function $\beta$ is dimensionless and is growing with the growth of $\frac{\omega}{\omega_0}$.

\noindent However, even for the models exhibiting the strongest wind input, the value of $\beta$ belongs to the interval $0 < \beta < 1$, for $\xi$ from the interval $0 < \xi < 3$. In some models (see, for example \cite{R15}) $\beta$ becomes negative for the waves propagating faster than the wind, or under large angle, with respect to the wind.

\noindent Looking at multiple experimental attempts to define $S_{in}$, one should notice that they need to be critically analyzed. The criticism is not about the integrity of measurements itself, but about the methodology used, the validity of data interpretation, and the possibility of transferring conclusions made in artificial environments to real ocean conditions.

\noindent A significant amount of the experiments, belonging to the so-called``fractional growth method" category, have been performed, through energy spectrum measurement in time, and calculation of the corresponding $\gamma$ through

\begin{eqnarray}
\label{FracGrowth}
\gamma(\omega, \phi) = \frac{1}{\varepsilon(\omega,\phi)} \frac{\partial \varepsilon(\omega, \phi)}{\partial t}
\end{eqnarray} 

\noindent Eq.(\ref{FracGrowth}) is, in fact, the linear part, or just two terms of the $HE$ Eq.(\ref{HE}). This method is intrinsically wrong, since it assumes that either advection $\frac{\partial \omega _{k} }{\partial \vec{k}} \frac{ \partial \varepsilon }{\partial \vec{r}}$ and nonlinear $S_{nl}$ terms of Eq.(\ref{HE}) are absent altogether, or relation 

\begin{eqnarray}
\label{LimFetch}
\frac{\partial \omega }{\partial \vec{k}} \frac{\partial \varepsilon }{\partial \vec{r}} =S_{nl}
\end{eqnarray} 

\noindent is fulfilled. 

\noindent The first assumption is simply not correct, since neglected terms are defining ocean conditions. 

\noindent The second assumption is almost fulfilled, indeed, since the sea is described by the $WTT$. But the terms in Eq.(\ref{LimFetch}) are large with respect to the terms in Eq.(\ref{FracGrowth}), and therefore there is no reason to neglect the terms of Eq.(\ref{LimFetch}). 

\noindent Regarding the ``fractional growth method", we are just citing the single publication by Plant \cite{R26} where, it seems, the author understood its scarcity.

\noindent As a matter of fact, it is natural to ask about the spectral correlation function $Q(\omega)$ between the surface elevation $\eta$ and the wind-induced pressure on the surface $P$:
\noindent
\begin{eqnarray}
\label{PressureCorr}
Q(\omega) = <\eta(\omega) P^*(\omega)> 
\end{eqnarray}
\noindent where brackets denote ensemble averaging, in Fourier space, and asterisk refers to the complex conjugate.

\noindent Unfortunately, the number of such experiments is limited, and not all of them have significant value for describing ocean phenomena. Also, one should not consider the experiments performed in laboratory conditions.

\noindent Consider, for example, the set of experiments described in \cite{R27}. These experiments were performed in the wave tank of $40 \,\, m$  length and $1 \,\, m$ depth. The wind was blowing at speeds up to $16 \,\, m/sec$, but they only studied short waves, no longer than $3 \,\, m$, moving no faster than $3.3 \,\, m/sec$.  Therefore, they studied the very short, wave-tail of the function $\beta$, in conditions far from those of the ocean. Another problem with flume data is upper physical confinement of the vertical velocity profile. The value of these measurements is not significant.

\noindent The same arguments apply to multiple precisely-performed measurements on the Lake George, Australia \cite{R28}. The depth of this lake is, on average, about $1\,\, m$. That is why the waves slower than $3.3 \,\, m/sec$ can propagate on its surface. The typical wind speed, corresponding to these measurements was $8-12 \,\, m/sec$. Therefore, while the results of these measurements are quite interesting, the obtained expression for $S_{in}$ is questionable, because it runs completely against quasi-linear theory. The quasi-linear theory predicts smoothing of the velocity profile $U(z)$, with the wave's development. The wind input growth rate, however, was increasing with the wave's energy level, in the experiments \cite{R29}.

\noindent After critically analyzing experiments on $S_{in}$ measurements, only three of them deserve attention. Those are the experiments by Snyder et al. \cite{R30}, Hsiao at al. \cite{R31} and Hasselmann at al. \cite{R32}. These experiments were performed in the open ocean and measured direct correlations of surface speed change and the pressure. The accuracy was not ideal and the data scatter was significant, presumably due to contemporary technologies. Therefore, their interpretation is quite ambiguous. The fact of this uncertainty was highlighted in the paper \cite{R15}. Either way, these experiments produced two well-known formulas for $\beta$. Next, we present $\beta$'s expressions, for the cases analyzed in the current paper.

\noindent For Snyder et al. \cite{R30}, Hasselmann-Bosenberg \cite{R32} case
\noindent
\begin{eqnarray}
\label{BetaSHB}
\beta_{SHB} = \left \{ \begin{array}{l} { 0.24 (\xi-1) \,\,\, for \,\,\, 1 < \xi < 4 } \\ 
{0, \,\,\, for \,\,\, \xi < 1 } \end{array}\right. 
\end{eqnarray} 

\noindent For Hsiao-Shemdin case \cite{R31}
\noindent
\begin{eqnarray}
\label{BetaHS}
\beta_{HS} = \left \{ \begin{array}{l} { 0.12 (0.85 \xi-1)^2 \,\,\, for \,\,\, \xi \geq 1 } \\ 
{0, \,\,\, for \,\,\, \xi < 1 } \end{array}\right. 
\end{eqnarray}

\noindent Let us notice that for $ZRP$ case \cite{R9}

\noindent
\begin{eqnarray}
\label{BetaZRP}
\beta_{ZRP} = 0.05 \xi^{4/3}
\end{eqnarray}

\noindent The differences between various $S_{in}$,  corresponding to Eqs.(\ref{BetaSHB})-(\ref{BetaZRP}), are significant. For many practical purposes, the spectral peak is located in the interval $1.5 < \xi < 2.5$, where the difference between Snyder and Hsiao-Shemdin functions is huge. Indeed

\noindent
\begin{eqnarray}  
&&\beta_S (1.5) = 0.12 \,\,\,\, \beta_{HS} (1.5) = 0.009 \\
&&\beta_S (2.5) = 0.36 \,\,\,\, \beta_{HS} (2.5) = 0.15
\end{eqnarray}

\noindent This serious difference is explained by lack of accuracy in both experiments (see \cite{R15}). Fig.6 of paper \cite{R30} and Fig.4 of the paper \cite{R31} show that the experimental data scatter has the same order as the mean values. Thus, the offered forms of the source functions Eqs.(\ref{BetaSHB}), (\ref{BetaHS}) are not seriously justified. However, Hsiao-Shemdin data appears to be more trustworthy. It seems quite obvious that the Snyder function overestimates the wind input, by several times. Presented numerical experiments justify this conjecture.

\noindent For the $ZRP$ function

\noindent
\begin{equation}
\beta_{ZRP}(1.5)=0.086 \,\,\,\,\,\, \beta_{ZRP}(2.5)=0.17
\end{equation}

\noindent and in the interval $1 < \xi < 3$  

\noindent
\begin{equation}
\beta_{SHB} (\xi) < \beta_{ZRP} (\xi) < \beta_S (\xi) 
\end{equation}

\begin{figure}
	\center
	\includegraphics[width=0.5\textwidth]{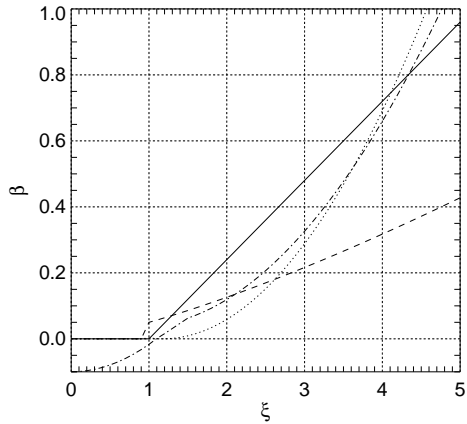}
	\caption{Four cases of function $\beta(\xi)$ along the wind ($\theta=0$) used in the numerical experiments. Solid line: Snyder-Hasselmann-Bosenberg case Eq.(\ref{BetaSHB}) , dashed line: Hsiao-Shemdin case Eq.(\ref{BetaHS}), dashed-dotted line: Chalikov case \cite{R25} and dotted line: ZRP case Eq.(\ref{BetaZRP})}. 
\label{FourBeta}
\end{figure}

\noindent Fig.\ref{FourBeta} presents one-dimensional plots of four functions, $\beta(\xi)$,  studied in numerical experiments presented below. We intentionally did not include the description of the sophisticated Chalikov  algorithm \cite{R25} for corresponding $\beta(\xi)$,  for the sake of space.

\noindent
\noindent We can conclude that, at the moment, there is no reliable parameterization of $S_{in}$, accepted by the entire oceanographic community. Keeping that fact in a mind, we decided to go our own way; this way is not focused on building new theoretical models, nor will it reconsider measurements of $S_{in}$.

\noindent For 68 years, since the works of Sverdrup and Munk  \cite{R53}, physical oceanography assimilated tremendous amount of wind-wave experimental data - wave energy and spectral peak frequency- as functions of limited fetch.  Such experiments are analyzed in works \cite{R5,R6,R7,R8}. On the other hand, numerical methods for solving the $HE$ Eq.(\ref{HE}), with exact $S_{nl}$ terms, have been improved significantly, for duration-limited and fetch-limited domains, as well.

\noindent Therefore, a new, purely pragmatic, approach for defining $S_{in}$ was proposed. The function $S_{in}$ has been chosen in such a way, so that numerical solutions of the Hasselmann equation explain the most known field experiments. The result was the $S_{in}$ function, described in detail in \cite{R9}, and named thereafter: the $ZRP$ function.

\noindent It is important to emphasize that work \cite {R9} assumed localization of energy dissipation, in short waves. This assumption contradicts widely accepted concepts, but we explain the differences in the following section.

\section{Two scenarios of wave-breaking dissipation term: spectral peak or high-frequency domination?} \label{TwoScenarios}

\noindent In this section, we explain why there is no need to use dissipation, in the spectral peak area.

\noindent Spectral peak frequency damping is a widely accepted practice, and is included as an option in the operational models $WAM$, $SWAN$ and $WaveWatchIII$. Notice that different operational models use completely different long-wave dissipation functions.

\noindent The form of $WAM3$ spectral peak dissipation used in this paper is given by the following definition \cite{R52} (the original notations are preserved):
\noindent
\begin{eqnarray}
\label{DissWAM}
S_{ds}(k,\theta) = C_{ds} \hat\sigma \frac{k}{\hat k} \left( \frac{\hat\alpha}{\hat \alpha_{PM}} \right)^2 N(k,\theta) \\ \nonumber
\hat \sigma = \left( \overline{\sigma^{-1}} \right)^{-1} \\  \nonumber
\hat \alpha = E {\hat \sigma}^4 g^{-2}
\end{eqnarray}
\noindent where $N(k,\theta)$ is the wave action spectrum, $\sigma$ is the frequency, $k$ is the wavenumber, $\theta$ is the angle, $E$ is the total energy, $C_{ds} = -2.36 \cdot 10^{-5}$, $\hat \alpha_{PM} = 3.02 \cdot 10^{-3}$ is the value of $\hat\alpha$ for $PM$ spectrum. Overline notation in $\left( \overline{\sigma^{-1}} \right)^{-1} $ means averaging over the spectrum.

\noindent This formula implicitely assumes that dissipation is concentrated in the long-wave region, and numerical experiments, below, show that it is indeed realized that way, see Fig.\ref{DissVsEnergy}.

\noindent It is important to emphasize that Eq.(\ref{DissWAM}) is not supported by laboratory or field experiments, nor by analytical theory, nor numerical simulations, in the framework of phase-resolving numerical models. This is a heuristic construction, and it is important to trace its origin. Eq.(\ref{DissWAM}) appeared in the paper of Komen et al. \cite{R54}, and exerted strong influences on future developments of physical oceanography.

\noindent The authors of article \cite{R54} analyzed the energy balance, in the surface of the wind-driven sea, and concluded that the introduction of artificial dissipation term Eqs.(\ref{DissWAM}) is necessary for explanation of experimental facts. This analysis was unsatisfactory for two reasons.

\noindent The authors of \cite{R54} considered that existence of the "fully developped" sea, shich is not only stationary in time, but homogeneous in space, as an obvious fact. For this reason they neglected the advection term $C_g \frac{\partial \varepsilon}{\partial x}$ in their analysis. In fact, all known stationary spectra vary with the fetch. It makes the concept of the "fully developped sea" doubtful.

\noindent Another weak point of the paper \cite{R54} is its uncritical use of the Snyder source function. As was shown before, this function has shaky foundations. Our numerical calculations show that it overestimates the energy growth rate by a factor of $5 \div 6$. The authors of Eqs.(\ref{DissWAM}) were using the Hsiao-Shemdin source function, which would hardly support Eqs.(\ref{DissWAM}) dissipation function, seeing in this case the balance had to be shifted to the dissipation side.

\noindent Anyways, for more than thirty years the dissipation term Eq.(\ref{DissWAM}), together with Snyder input term, dominated in the operating models. These choices became a sort of credo, in physical oceanography. The purpose of the presented paper is its revision.

\noindent The dissipation of water surface waves, due to white-capping, is an extremely important physical phenomenon, not yet properly studied. 
M.Longuet-Higgins spent a lot of effort to develop an analytical theory of the wave-breaking \cite{R82,R83,R84,R85}. He found  the set of interesting exact and approximate solutions of the Euler equations, describing potential flow, of ideal fluid with the free surface, but didn't solve the problem completely.

\noindent The difficulty of development in wave-breaking analytical theory is explained by sophisticated mathematical reasoning. Most probably, the system of Euler equations, for incompressible ideal fluid potential flow, with free surface on deep water in $1+1$ geometry (i.e. depth coordinate and one horizontal coordinate) is the completely integrable system. It has too many peculiar features: cancellation of non-trivial four-wave interactions \cite{R81}, presence of an indefinite number of extra motion constants \cite{R87}, partial solutions describing propagating capillary waves, expressed in elementary functions (the ``Crapper solution" \cite{R103}). So far, the exact integrability, for the general time-dependent problem, is established in the exotic case of ``asymptotically upweiling flow", in the absence of gravity \cite{R88}. An infinite number of exact solutions were found in the paper. Some of those solutions were published long ago by Longuet-Higgins \cite{R85}.

\noindent Integrability makes the theory of white-capping complicated for the following reason. Integrability means absence of a universal scenario, of this effect. From the view-point of general nonlinear wave dynamics, wave-breaking is an example of ``weak-collapse" \cite{R86}. Such collapses are described, as a rule, by self-similar solutions.

\noindent Breakers, described by self-similar solutions

\begin{eqnarray}
\label{SelfSimilar}
\eta (x,t)=g(t_0-t)^{2} F\left(\frac{x-x_0}{g(t_0-t)^{2} } \right)
\end{eqnarray}
 
\noindent were studied analytically and numerically, in the framework of the simplified (and non-integrable!) $MMT$ (Maida-McLaughlin-Tabak) model of Euler equations \cite{R33}. Here $\eta(x,t)$ is the water surface elevation, $x$ and $t$ are spatial coordinate and time, respectively, and $F$ is self-similar function.

\noindent Solution Eq.(\ref{SelfSimilar}) describes formation of the wedge, with the top at $x=x_0$, at the moment of time $t=t_0$. Exact Euler equations have similar solutions, describing formation of locally stationary ``corner flow solution", with angle $120^o$ (see Longuet-Higgins \cite{R83}). But in the $MMT$ model, the non-integrable case, the self-similar breaker is a ``global attractor". In other words, all breakers are self-similar. The exact Euler equation has the same self-similar solution describing formation of the ``Stokes corner flow" \cite{R83}, but now it is not a generic scenario. In the general case, formation of the wedge is only the first stage of the breaker evolution. Later it ejects an inclined ``Dirichlet jet" \cite{R82}, which plunges back into the water and transforms mechanical energy to heat. This scenario was qualitatively described in an article of Longuet-Higgince, and is supported by many numerical simulations  \cite{R89}-\cite{R92}, laboratory experiments \cite{R104}, and field observations \cite{R92}-\cite{R96}. The literature on this subject is huge, and only a small portion of it is cited.

\noindent In spite of the complexity of this scenario, in terms of Fourier transform, the physical picture of the phenomenon is, more or less, universal. On the stage of wedge formation, the spatial Fourier spectrum, of energy, forms a ``fat tail". Up to a certain moment of time, this spectrum is reversible in time. Plunging of the jets causes formation of the drops and bubbles, leading to dissipation of the energy and irreversibility. This is the mechanism of ``high-frequency dissipation". The presence of high-frequency dissipation ``chops off" the end of the tail, and violates the tail invertability. Low and high harmonics, however, are strongly coupled in this event, due to strong, nonlinear, non-local, interaction, and deformed high wave-numbers, so the tail, almost immediately, returns to the spectral peak area. As soon as  the fat spectral tail returns to the spectral peak area, total energy in the spectrum diminishes, causing settling of the spectral peak at a lower level of energy. This process of ``shooting" of the spectral tail toward high wave-numbers, and its returning back, due to wave breaking, is the real reason of ``sagging down" of the energy profile in the spectral peak area. This was erroneously associated with the presence of the damping in the spectral peak area. This explanation suggests that individual wave-breaking studies \cite{R29,R34} do not prove the presence of spectral peak damping.

\noindent There is another question of fundamental importance.  What is the speed, $C_b$, of breaker propagation? This $C_b$ is connected with the characteristic length of a breaker by $P_b \simeq \frac{C_b^2}{g}$. The breaker propagation speed is the the subject of direct measurements. Breakers produce strips of foam, and propagation of these strips can be traced relatively easily. The results of numerous experiments performed by Huang et al. \cite{R35,R36,R37,R38}, Gemmrich et al. \cite{R97} gave, approximately, the same result: most of the breakers are ``slow". Their propagation speed is $C_b \simeq 0.2 C_p$. Slow breakers are quasi-one-dimensional, but they propagate in a broad sector of angles with respect to the wind. Some breakers are fast ( $C_b \simeq C_p$ ). Fast breakers propagate in  the same direction as the leading wave. What is 
important is that ``slow" and ``fast" breakers are formed, but for completely different reasons. Let us look at the $KZ$ spectrum 

\begin{equation}
\varepsilon_\omega \simeq \omega^{-4},\,\,\,\,\,\,\,I_k \simeq k^{-5/2}
\end{equation}

\noindent This spectrum is concentrated on fractals, non-smooth functions, and cannot be extended very far in the high frequency. At $\omega \simeq 3 \omega_p$ the $KZ$ spectrum turns into the Phillips spectrum $\varepsilon \simeq \omega^{-5}, \,\,\, I_k \simeq k^{-3}$, which is concentrated on wedge-like functions. Thus, formation of ``slow breakers" is an unavoidable consequence of the energy flux to high frequency region, provided by four-wave, nonlinear, wave interaction. Thus, it is reasonable to suppose that wave-breaking dissipation is localized in short scales.

\noindent The population of waves having frequencies $3 \div 4$ times bigger than spectral maximum frequency is called ``Phillips sea" \cite{R58,R59,R60}, which we call for the sake of brevity the ``short waves". The ``Phillips sea" contains no more than $2\%$ of the total wave energy, but the whole energy dissipation, fueled by energy flux from long waves, is happening right there. It is proved, experimentally, that ``Phillips sea" is described by a universal Phillips spectrum $\varepsilon \simeq \frac{\alpha g^2}{\omega^5}$, where $\alpha \simeq 0.01$ is a dimensionless constant, while for the Pierson-Moscowitz spectrum $\alpha = 0.0081$ \cite{R58,R59,R60}.

\noindent ``Phillips sea" is quite an interesting physical object. It contains breakers of different, statistically uniformly distributed, sizes \cite{R60}, down to characteristic wave length $\lambda \simeq 3 \div 5$ cm, where capillary effects become essential.

\noindent The exact form of the ``Phillips sea" energy dissipation function is unknown. Recently, a quite plausible model of such function has been presented in \cite{R63}, which is hoped to become the subject of oceanographic community discussion. 

\noindent What's about the ``fast breakers"? They rarely appear in $1+1$ geometry, where nontrivial four-wave interactions are canceled out, and there is no energy flux to high wavenumbers.

\noindent The steepness, in the conditions of typically developed wave turbulence, is not big: $\mu = <\nabla \eta^2 >^{1/2} \sim 0.1$, or even smaller. Because this value is very far from limiting steepness of Stokes wave $\mu \simeq 0.3$, these waves are, essentially, weakly-nonlinear. Besides those waves, shorter waves inevitably develop, having steepness approaching the critical one, and those waves break.

\noindent However, there is also another process- the modulational instability or ``wave grouping"- which leads to spatial inhomogeneity and formation of ``rough waves" propagating, with the spectral peak velocity. Theory of rogue wave formation is a separate, and interesting phenomenon, which is discussed in many articles (see, for instance \cite{R98,R99,R100}), but is not the subject of the current paper.

\noindent It is important that direct numerical solutions of both exact \cite{R56} and approximate \cite{R105} primordial Euler equations show that dissipation of the rogue waves does not make any significant contribution into energy balance of wind-driven seas. Thus, the main conclusion about the dissipation taking place in short scales remains unchanged.

\section{Numerical experiments set-up} \label{NumSetUp}

\noindent The subject of numerical simulation was the stationary $HE$ Eq.(\ref{HEstat}), for different wind input functions. A total of 5 different wind inputs have been tested in the frame of stationary $HE$:

\noindent 
\begin{equation}
\label{HEsimplified}
\frac{\partial\omega_k}{\partial \vec{k}} \frac{\partial \varepsilon}{\partial \vec{r}} = S_{nl} + S_{in} + S_{diss} 
\end{equation}

\noindent The first 4 tests were done assuming the absence of low-frequency dissipation and the presence of ``implicit" high-frequency dissipation. 

\noindent All simulations were performed with the help of the $WRT$ method \cite{R68,R114,R115}, previously used in \cite{R46,R47,R48,R50,R63,R69,R72}, on the grid of $71$ points in frequency and $36$ points in angle domains. A constant wind of speed $10 \,\, m/sec $ is assumed to be blowing away from the shore line, along the fetch. The assumption of the constant wind speed is a necessary simplification, due to the fact that the numerical simulation is being compared to various data from field experiments, and the considered set-up is the simplest physical situation, which can be modeled.

\subsection{The details of ``implicit" damping implementation}

\noindent One should specifically stop and note details of the ``implicit" high-frequency damping, used in all five numerical simulations. Including the ``implicit" damping consists in continuation of the spectral tail by Phillips law \cite{R58} $A(\omega_0) \cdot  \omega^{-5}$, where $A(\omega_0)$ is the parameter dynamically changing in time.
 
\noindent The coefficient $A(\omega_0)$, in front of $\omega ^{-5} $, is not exactly known, but is not required to be defined in explicit form - it is dynamically determined from the continuity condition of the spectrum, at frequency $\omega_0$, on every time step. In other words, the starting point of the Phillips spectrum coincides with the last point  of the dynamically changing spectrum, at the frequency point $\omega_{crit} = 2 \pi f_{crit}$, where $f_{crit} \simeq 1.1\,\,Hz$, as per Resio and Long experimental observations \cite{R50}. This is the way the high frequency ``implicit" damping is incorporated into the alternative computational framework of $HE$. 

\noindent One should note recently developed analytical models, which automatically describe the transition from  the $KZ$ spectrum $\omega^{-4}$ to Phillips tail $\omega^{-5}$  \cite{R63}. Such modification of the ``implicit" damping is in future plans, but the question of the finer details of the high-frequency ``implicit" damping structure is of secondary importance, at the current ``proof of the concept" stage, of the alternative framework development.

\subsection{$WTT$ facts used in numerical simulation}

\noindent As a rule, confirmed by field and numerical observations, the wave energy spectrum has sufficiently sharp peak at $\omega \simeq \omega_p$. However, almost immediately after the spectral peak at $\omega \simeq 1.5  \omega_p$ the advection term $\frac{\partial\omega}{\partial \vec{k}} \frac{\partial \varepsilon}{\partial \vec{r}}$ becomes insignificant, and the original stationary $HE$ Eq.(\ref{HEstat}) is transformed into
\noindent
\begin{equation}
\label{StatEq}
S_{nl} + S_{in} + S_{diss} = 0
\end{equation}

\noindent Comparison with Eq.({\ref{HEsplit}}) shows that Eq.(\ref{StatEq}) can be rewritten in the form 

\noindent
\begin{equation}
\label{}
\varepsilon(\omega,\theta) = \frac{F(\omega,\theta)}{\Gamma(\omega,\theta)-\gamma(\omega,\theta)+\gamma_{diss}(\omega,\theta)} 
\end{equation}

\noindent As it was shown in \cite{R1,R2}, nonlinear dissipation $\Gamma(\omega,\theta)$ in the ``universal area" $\omega > 1.5 \omega_p$
is several times greater than wind forcing term. Therefore, Eq.({\ref{StatEq}}) can be rewritten, as a first approximation, by
\noindent
\begin{equation}
\label{SnlZeroAnother}
S_{nl}(\omega,\theta) = 0
\end{equation}

\noindent or

\begin{equation}
\varepsilon(\omega,\theta) = \frac{F}{\Gamma}
\end{equation}

\noindent As was mentioned before, this equation has a rich family of solutions. The simplest and best known solution is the isotropic Zakharov-Filonenko solution \cite{R41}
\noindent
\begin{equation}
\label{KZsolution}
\varepsilon(\omega,\theta) = \frac{C_p g^{4/3}}{\omega^4} P^{1/3} = \frac{\beta_{KZ}}{\omega^4}
\end{equation}

\noindent Here $P$ is the energy flux into the high wavenumbers region. 

\noindent The energy density flux per square unit, in the atmosphere, is $P_{sq} = \rho_a U^3$, where $\rho_a$ is the atmosphere density and $U$ is the wind speed. For $U=10\,\,m/sec$, $P_{sq}\simeq 1.2 \,\,kW$. A relatively small part of this flux is transferred to the water.  According to Hwang and co-authors \cite{R37,R38}, the estimated amount, transferred to the ocean, is $P_0 \simeq 0.1\,\, Wt$. Approximately one third of this amount is spent into energy flux formation, toward high wave numbers. In ``oceanographic" normalization this flux has to be divided by $\rho_w g$, where $\rho_w\simeq 10^3\,\,kg/m^3$. Finally, the energy flux toward small scales is 

\noindent
\begin{equation}
P\sim 2 \div 3 \cdot 10^{-6}\,\,m^2/sec
\end{equation}

\noindent This expression agrees with the presented numerical experiments. The Kolmogorov constant in the Eq.(\ref{KZsolution}) can be found numerically \cite{R1}. Recently, its value has been found more rigorously \cite{R78}:

\noindent
\begin{equation}
C_p = 4 \pi \cdot 0.194 = 2.43
\end{equation}

\noindent One can estimate the characteristic value of $\beta_{KZ}$:

\noindent
\begin{equation}
\label{BetaKZ}
\beta_{KZ} = C_p g^{4/3} P^{1/3} \simeq 0.6\,\,m^2/sec^3
\end{equation}

\noindent According to $WTT$, the value of $\beta_{KZ}$ should be constant, somewhere in the region $1.5 \omega_p < \omega < \omega_{crit}$. Here, $\omega_{crit} = 2 \pi \cdot 1.1 = 6.91$ is the critical frequency at which the ``implicit" damping is turning on. The energy flux, for this area, is diminishing proportionally to $\omega^{-3}$, and $\beta_{KZ}$ is not  constant anymore - one has to substitute $\beta_{KZ}$ by $\beta_{KZ} \cdot \frac{\omega_{crit}}{\omega}$.

\noindent The general, anisotropic $KZ$ solution, with zero wave-action flux from $\omega \rightarrow \infty $, can be presented in the form \cite{R1}

\begin{equation}
\label{GeneralKZ}
\varepsilon(\omega, \theta) = \frac{\beta_{KZ}}{\omega^4} F \left(\frac{\omega_s}{\omega},\theta \right)
\end{equation}

\noindent where $\omega_s=\frac{M}{P}$, $M$ - the momentum flux to the small scale region. The frequency $\omega_s$ depends on the shape of $S_{in}$, in a typical case $\omega_s \simeq \omega_p$. $F\left( \frac{\omega_s}{\omega}, \theta \right)$ is a ``structural function". It is established that in the limit $\frac{\omega_s}{\omega} \rightarrow 0$ (see Katz et al. \cite{R107}, Zakharov \cite{R1}): 

\begin{equation}
\label{KKZ}
F\left( \frac{\omega_s}{\omega}\right) \rightarrow 1 + c_2 \frac{\omega_s}{\omega} \cos \theta
\end{equation}

\noindent where $c_2$ is the ``second Kolmogorov constant".

\noindent If one uses, for $S_{nl}$, the ``diffusion approximation" \cite{R106}, the structural function is known

\begin{equation}
F\left( \frac{\omega_s}{\omega}, \theta \right) = \left( 1 + \frac{\omega_s}{\omega} \cos \theta \right)^{1/3}
\end{equation}

\noindent As far as $F\rightarrow 1$ at $\omega \rightarrow \infty$, the $KZ$ solution Eq.(\ref{GeneralKZ}) describes well-known ``angular spreading". This solution becomes isotropic at $\omega \rightarrow \infty$.

\noindent One can introduce

\begin{equation}
f\left( \frac{\omega_s}{\omega} \right) = \frac{1}{2\pi} \int_0^{2\pi} F \left( \frac{\omega_s}{\omega}, \theta \right) d\theta
\end{equation}

\noindent From Eq.(\ref{KKZ}), one can see that $f\left( \frac{\omega_s}{\omega}\right) \rightarrow + \lambda \left(\frac{\omega_s}{\omega} \right)^2$ + ...  

\noindent A more detailed study, of the functions $F\left( \frac{\omega_s}{\omega}, \theta \right)$ and $f\left( \frac{\omega_s}{\omega} \right)$, is an urgent theoretical problem, but is out of the scope of the current paper. One can expect, however, that $f\left( \frac{\omega_s}{\omega} \right)$, for $\omega > 2 \omega_p$, is close to $1$. Presented calculations confirm this conjecture: the compensated angle-averaged spectrum $<\epsilon> \omega^4$ is constant, up to $20\%$ accuracy inside, the spectral band $0.4 Hz < f < 1 Hz$.

\noindent Notice that the average energy spectrum, $<\varepsilon>$, decays in this spectral band, by a factor of $40$, and the difference of the $KZ$ spectrum and Phillips spectrum is essential.

\noindent Let's discuss self-similar solutions of the conservative $HE$ Eq.(\ref{LimFetch}). This equation has a family of self-similar solutions, which can be conveniently rewritten in the form
 
\noindent
\begin{eqnarray}
\label{SSS}
\varepsilon(\omega,\theta,\chi) &=& \chi^{p+q} F(\xi, \theta) \\
\xi &=& \omega \chi^q
\end{eqnarray} 

\noindent where $F(\chi,\theta)$ is the function, satisfying the relation 

\noindent
\begin{eqnarray}
\frac{\cos \theta}{2 \xi}\left[ (p+q) F + q \xi \frac{\partial F}{\partial \xi} \right] = S_{nl}
\end{eqnarray} 

\noindent and $q$ and $p$ are the constants, connected by the ``magic relation" Eq.(\ref{MagicRelation}). If the self-similar solution is realized, then dimensionless energy and frequency are the power functions, of the dimensionless fetch; see Eq.(\ref{Peq}), (\ref{Qeq}).

\noindent It was shown in \cite{R9} that self-similar solutions also exist for the case of wind input, if $\beta(\xi)$ in Eq.(\ref{GammaDefinition}) is the power function of the frequency:

\noindent
\begin{equation}
\beta = \omega^s f(\theta)
\end{equation}
  
\noindent The constants $p$ and $q$ are defined now, unambiguously as 

\noindent
\begin{eqnarray}
q &=& \frac{1}{2+s} \\
p &=& \frac{8-s}{2(2+s)}
\end{eqnarray}

\noindent As was already mentioned, practically all ocean field measurements demonstrate power dependencies Eq.(\ref{Peq}), (\ref{Qeq}). However, there is scattering in the definition of the exponent $p$, in the range $0.74 < p < 1.1$. It's quite possible that this scattering is due to absence of the universal expression for $S_{in}$, suitable for any atmospheric boundary layer state.

\noindent The experiments of Kahma (see \cite{R108} for references), performed for ``stable" and ``unstable" atmosphere, gave different values of $p$. However, the ``magic relation" Eq.(\ref{MagicRelation}) still holds true for those different cases.
This fact holds the promise that $WTT$ always works.

\noindent More than the half of the numerical experiments have the values $p=1$, $q=0.3$. Such self-similarity occurs if $s=4/3$. This fact, together with field experimental data \cite{R50},  leads to the appearance of the $ZRP$ wind input term \cite{R9}.

\section{Numerical study of different wind input models} \label{CheckingFramework}

\noindent The current section presents the results of numerical simulation, of different wind input models, in the alternative framework of $HE$, for limited fetch statement, based on the following:

\begin{enumerate}
\item Exact $S_{nl}$ term
\item Absence of spectral maximum dissipation (excluding special $WAM3$ case)
\item High-frequency ``implicit" dissipation 
\end{enumerate} 

The first numerically studied wind input model is described in the previous section's $ZRP$ model:

\noindent
\begin{eqnarray}
&&S_{in}(\omega,\phi) = \gamma(\omega,\phi)\cdot \varepsilon(\omega,\phi) \\
&&\gamma = 0.05 \frac{\rho _{air} }{\rho _{water} } \omega \left(\frac{\omega }{\omega _{0} } \right)^{4/3} f(\theta ) \label{ZRP1} \\
&&f(\theta ) = \left\{\begin{array}{l} {\cos ^{2} \theta {\rm \; \; for\; -}\pi {\rm /2}\le \theta \le \pi {\rm /2}} \\ {0{\rm \; \; otherwise}} \end{array}\right. \\
&&\omega _{0} = \frac{g}{u_{10} }, \,\,\,  \frac{\rho _{air} }{\rho _{water} } =1.3\cdot 10^{-3}
\end{eqnarray}

\begin{figure}
        \centering
        \begin{subfigure}[b]{0.45\textwidth}
                \includegraphics[width=\textwidth]{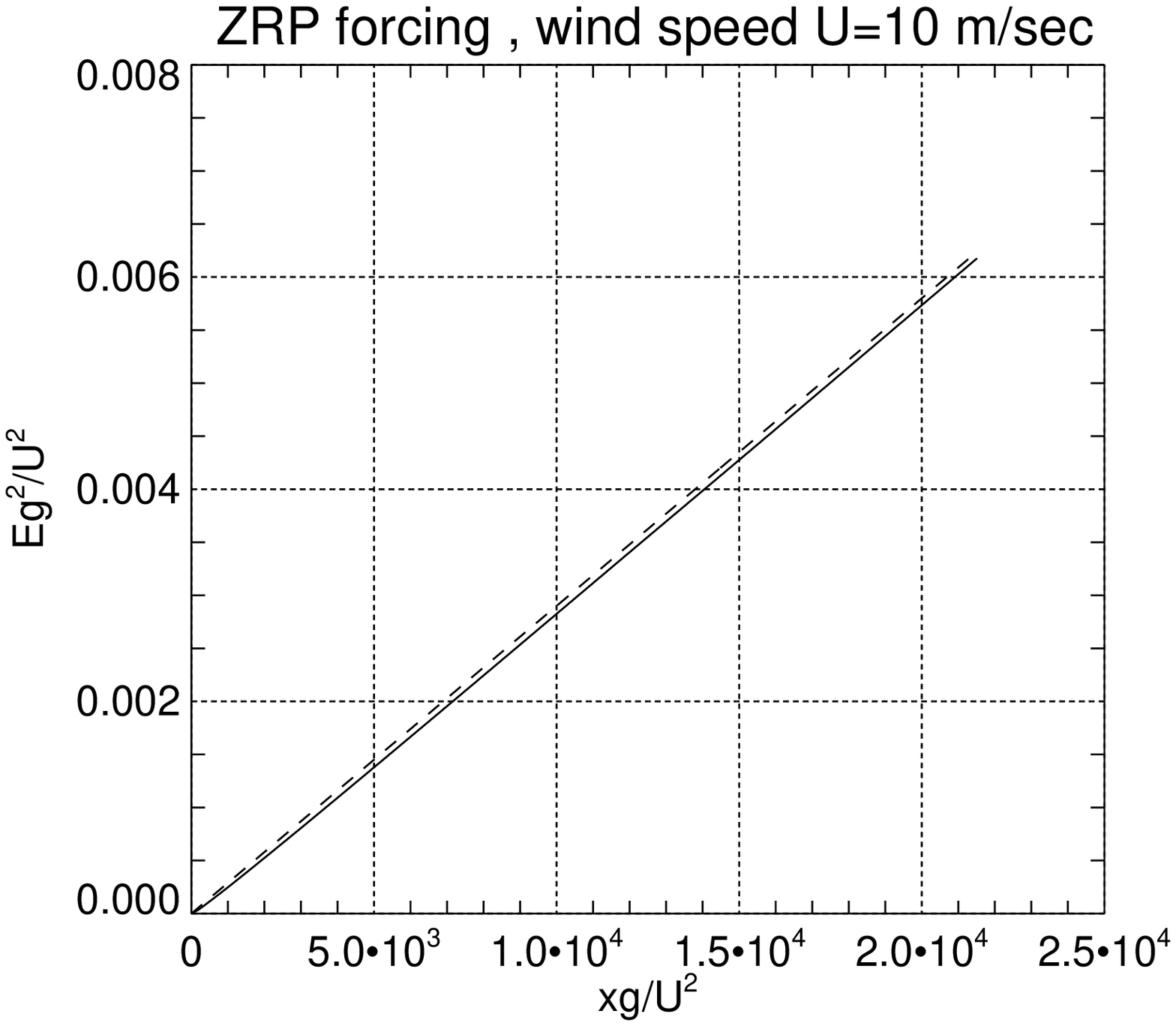}
                \caption{Dimensionless energy dependence on the dimensionless fetch, in numerical experiment. Dashed line -- theoretical fit by $2.9\cdot 10^{-7} \cdot \frac{xg}{U^{2}}$ }
                \label{AltFormA}
        \end{subfigure}
\qquad
%(or a blank line to force the sub-figure onto a new line)
        \begin{subfigure}[b]{0.45\textwidth}
                \includegraphics[width=\textwidth]{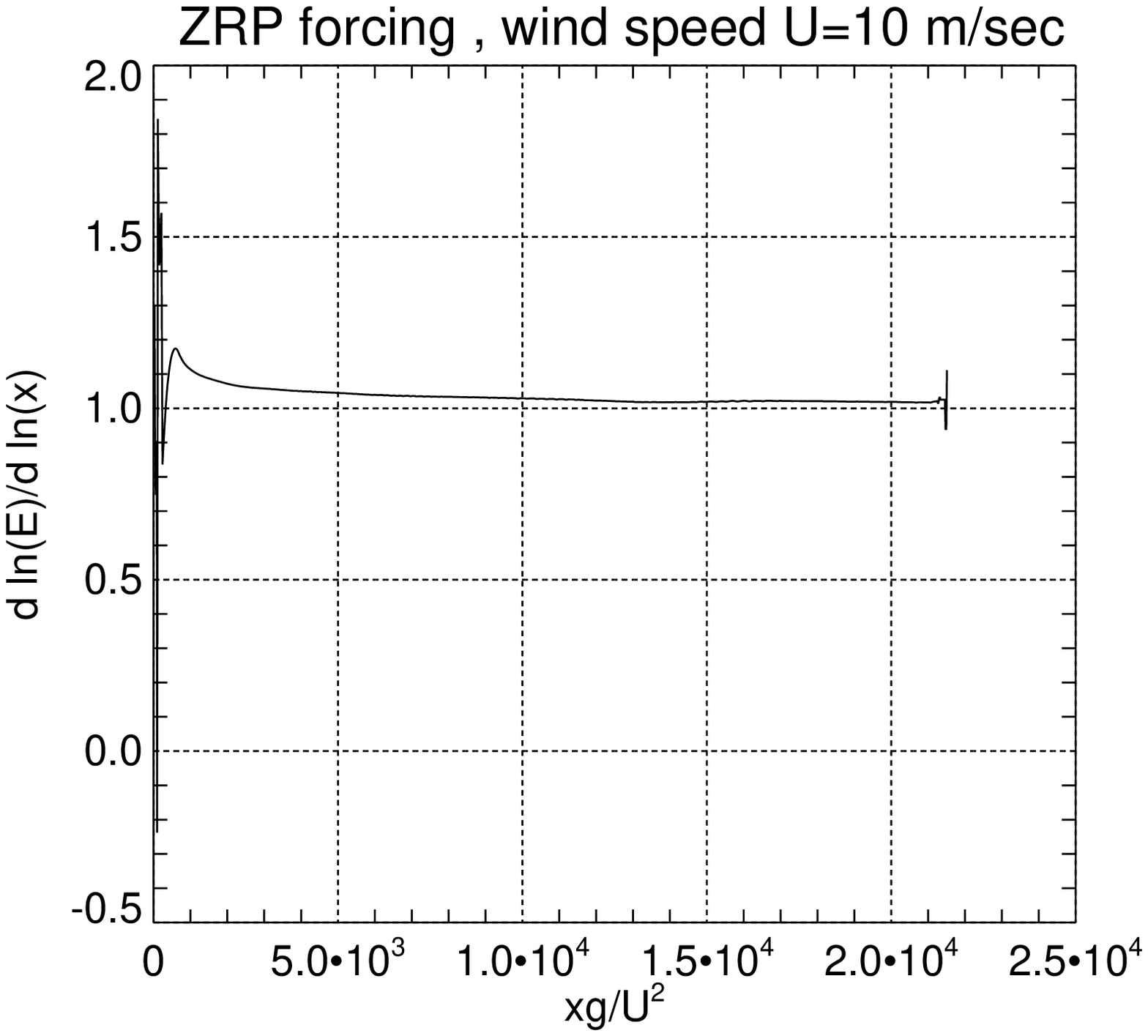}
                \caption{Local exponent $p$, of the energy growth, as a function of dimensionless fetch.}
                \label{AltFormB}
        \end{subfigure}
      \caption{}\label{AltForm}
\end{figure}

\noindent Fig.\ref{AltFormA} shows total energy growing along the fetch, by the power law
 
\noindent
\begin{eqnarray}
\hat\varepsilon = \varepsilon_0 \chi \nonumber
\end{eqnarray}

\noindent in accordance with Eq.(\ref{Peq}) for $p=1.0$, see Fig.\ref{AltFormB}.

%%%%%%%%%%%%%%%%%%%%%%%%%%%%%%%%%%%%%%%%%%%%%%%%%%%%%%%%%%%%%%%%%%%%%%%%%%%%%%%%%%%%%%%%%%%%%%%%%%

%%%%%%%%%%%%%%%%%%%%%%%%%%%%%%%%%%%%%%%%%%%%%%%%%%%%%%%%%%%%%%%%%%%%%%%%%%%%%%%%%%%%%%%%%%%%%%%%%%

\begin{figure}
        \centering
        \begin{subfigure}[b]{0.45\textwidth}
                \includegraphics[width=\textwidth]{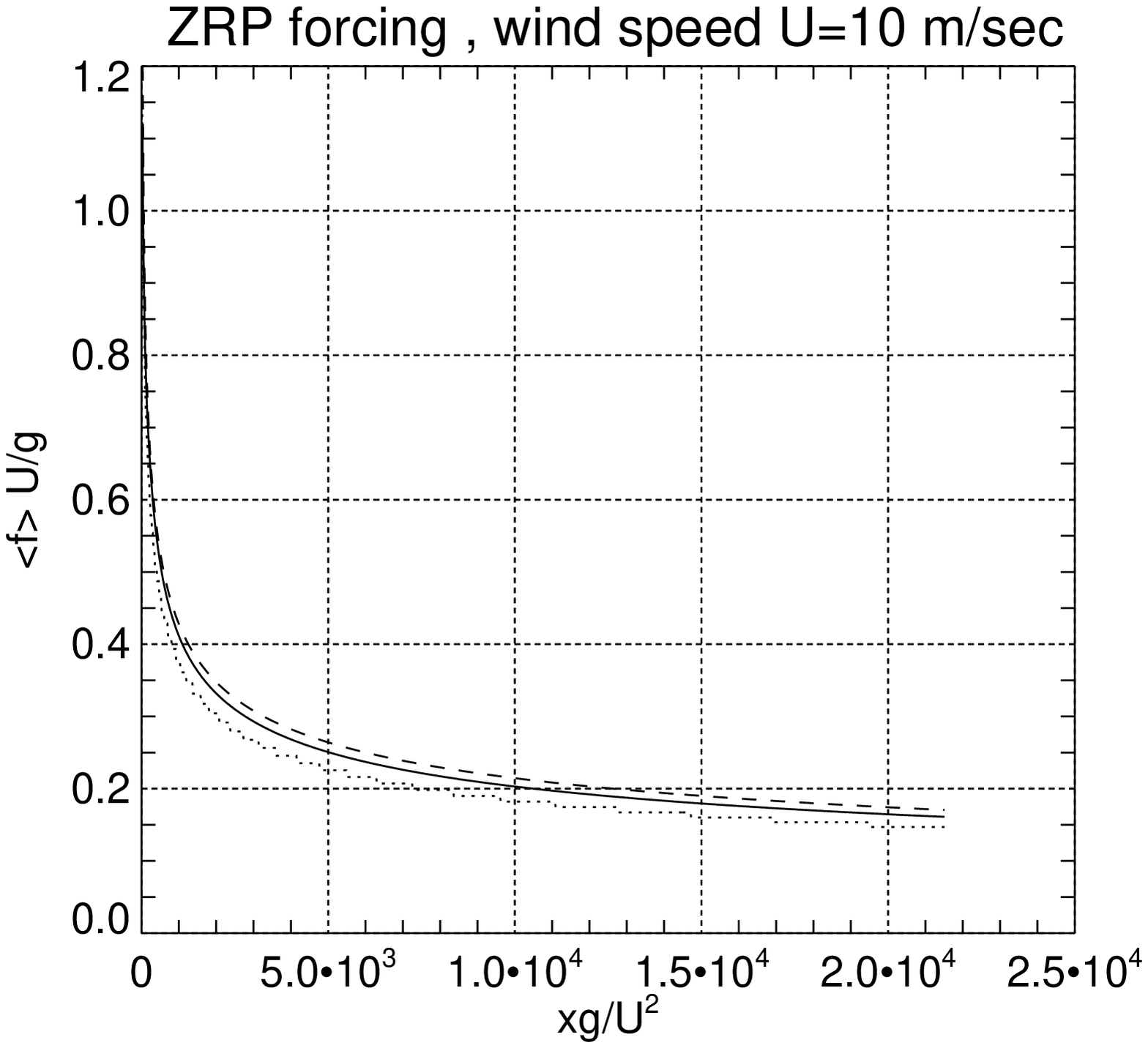}
                \caption{Dimensionless mean frequency, as a function of dimensionless fetch (solid line), calculated as $<f>=\frac{1}{2\pi}\frac{\int\omega n d\omega d\theta}{\int n d\omega d\theta}$, where $n(\omega,\theta)=\frac{\varepsilon(\omega,\theta)}{\omega}$ is the wave action spectrum. The dotted line is the peak frequency $f_p=\frac{\omega_p}{2\pi}$, and the dashed is the theoretical fit $3.4\cdot \left(\frac{xg}{U^{2}} \right)^{-0.3}$. }
                \label{MeanFreqA}
        \end{subfigure}
\qquad
%(or a blank line to force the sub-figure onto a new line)
        \begin{subfigure}[b]{0.45\textwidth}
                \includegraphics[width=\textwidth]{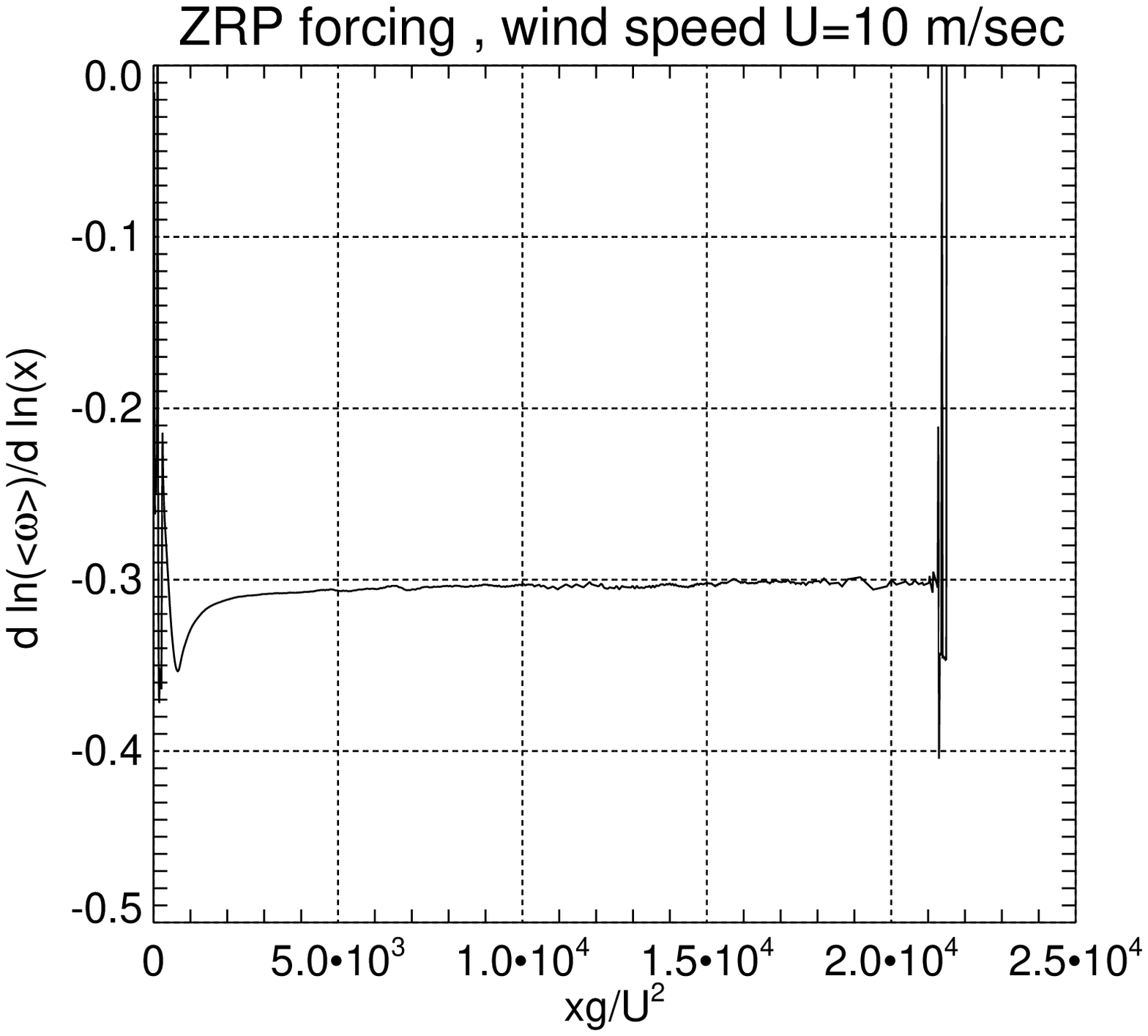}
                \caption{Local exponent $-q$ of the mean frequency as the function of dimensionless fetch.}
                \label{MeanFreqB}
        \end{subfigure}
      \caption{}\label{MeanFreq}
\end{figure}

\noindent Dependence of mean frequency, on the fetch, shown in Fig.\ref{MeanFreqA}, demonstrates the law

\noindent
\begin{eqnarray}
\hat\omega = \omega_0 \chi^{-0.3} \nonumber
\end{eqnarray}

\noindent in good correspondence with self-similar dependence Eq.(\ref{Qeq}), for $q=0.3$, see Fig.\ref{MeanFreqB}. 

\noindent Fig.\ref{MeanFreqA} presents, not only mean frequency, but also the maximum spectral frequency. Their difference, however, is so small, that we will not distinguish between them, hereafter.

%%%%%%%%%%%%%%%%%%%%%%%%%%%%%%%%%%%%%%%%%%%%%%%%%%%%%%%%%%%%%%%%%%%%%%%%%%%%%%%%%%%%%%%%%%%%%%%%%%

\begin{figure}
        \centering
                \includegraphics[width=0.5\textwidth]{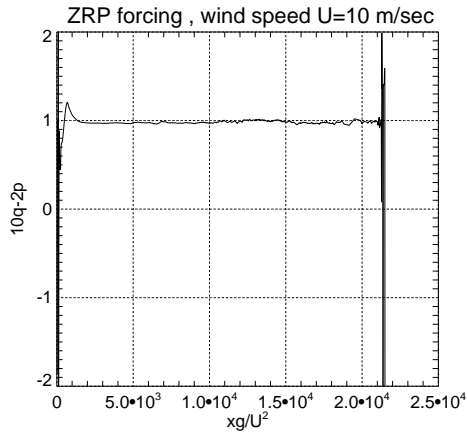}
                \caption{ ``Magic relation" $(10q-2p)$ as the function of dimensionless fetch for $ZRP$ wind input term.}
	\label{MagicNumber}
\end{figure}

\noindent The result of the ``magic relation" check is presented in Fig.\ref{MagicNumber}.
It presents the relation as a function of the fetch. It strongly agrees with the self-similar prediction of Eq.(\ref{MagicRelation}).

%%%%%%%%%%%%%%%%%%%%%%%%%%%%%%%%%%%%%%%%%%%%%%%%%%%%%%%%%%%%%%%%%%%%%%%%%%%%%%%%%%%%%%%%%%%%%%%%%%

%%%%%%%%%%%%%%%%%%%%%%%%%%%%%%%%%%%%%%%%%%%%%%%%%%%%%%%%%%%%%%%%%%%%%%%%%%%%%%%%%%%%%%%%%%%%%%%%%%

\noindent Table \ref{Table1} presents the results \cite{R6} of calculating the exponents $p$ and $q$ (see Eqs.(\ref{Peq}), (\ref{Qeq})), for 12 different experimental observations, with the last row corresponding to a limited fetch growth numerical experiment, within the alternative $ZRP$ framework. The value of $C=10q-2p$, averaged over the experiments, is $<C>=0.95$. One can see correspondence with the predicted, theoretical, value $C_t = 1$, as well as the numerical result. One should note interpretations of the $JONSWAP$ experiment, by different experts, provided different values of $p$ and $q$, and, correspondingly, $C$. 

\noindent 

\begin{table}
\centering
\begin{tabular}{|p{2.8in}|p{0.2in}|p{0.2in}|p{0.8in}|} \hline 
\textbf{Experiment} & $p$ & $q$ & $C=10q-2p$ \\ \hline 
Babanin, Soloviev 1998 Black Sea & 0.89 & 0.28 & 1.02 \\ \hline 
Walsh et al. (1989) US coast & 1.0 & 0.29 & 0.90 \\ \hline 
Kahma, Calkoen (1992) unstable & 0.94 & 0.28 & 0.92 \\ \hline 
Kahma, Pettersson (1994) & 0.93 & 0.28 & 0.94 \\ \hline 
JONSWAP by Davidan (1980)  & 1.0 & 0.28 & 0.80 \\ \hline 
JONSWAP by Phillips (1977)  & 1.0 & 0.25 & 0.75\\ \hline 
Kahma, Calkoen (1992) composite  & 0.9 & 0.27 & 0.90 \\ \hline 
Kahma (1981, 1986) rapid growth  & 1.0 & 0.33 & 1.03 \\ \hline 
Kahma (1986) average growth  & 1.0 & 0.33 & 1.03 \\ \hline 
Donelan \textit{et al. }(1992) St Claire  & 1.0 & 0.33 & 1.03 \\ \hline 
JONSWAP by Hasselmann et al. (1973)  & 1.0 & 0.33 & 1.03 \\ \hline 
Mitsuyasu et al. (1971)  & 1.0 & 0.33 & 1.03 \\ \hline 
ZRP numerics & 1.0 & 0.3 & 1.00 \\ \hline 
\end{tabular}
\caption{Exponents $p$ and $q$ (see Eqs.(\ref{Peq}),(\ref{Qeq})) for 12 different experimental observations \cite{R6} with the last row corresponding to limited fetch growth numerical experiment within alternative $ZRP$ framework.}
\label{Table1}
\end{table}

\begin{figure}
        \centering
	\includegraphics[width=0.5\textwidth]{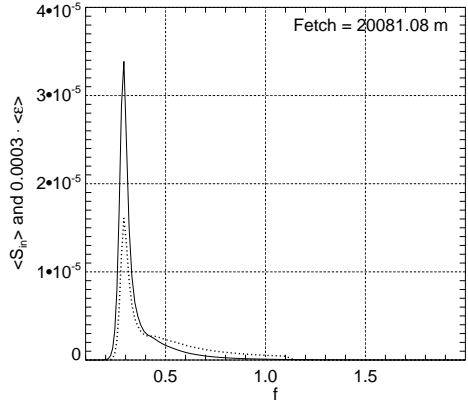}
	\caption{ Typical, angle averaged, wind input function density $<S_{in}> = \frac{1}{2 \pi}\int \gamma(\omega,\theta) \varepsilon(\omega,\theta) d\theta$ and angle averaged spectrum $<\varepsilon> = \frac{1}{2\pi} \int \varepsilon(\omega,\theta) d\theta$ (solid line) as the functions of the frequency $f=\frac{\omega }{2\pi}$.}
	\label{SpecInLinCoord}
\end{figure}

\noindent Let's proceed with the analysis of numerical spectra. Typical, angle averaged, wind input function density $<S_{in}>$ and angle averaged spectrum, in linear coordinates, are presented on Fig.\ref{SpecInLinCoord}. It is seen that a major portion of the wind forcing is concentrated in the spectral peak vicinity. 

%%%%%%%%%%%%%%%%%%%%%%%%%%%%%%%%%%%%%%%%%%%%%%%%%%%%%%%%%%%%%%%%%%%%%%%%%%%%%%%%%%%%%%%%%%%%%%%%%%

%%%%%%%%%%%%%%%%%%%%%%%%%%%%%%%%%%%%%%%%%%%%%%%%%%%%%%%%%%%%%%%%%%%%%%%%%%%%%%%%%%%%%%%%%%%%%%%%%%

\begin{figure}
        \centering
	\includegraphics[width=0.5\textwidth]{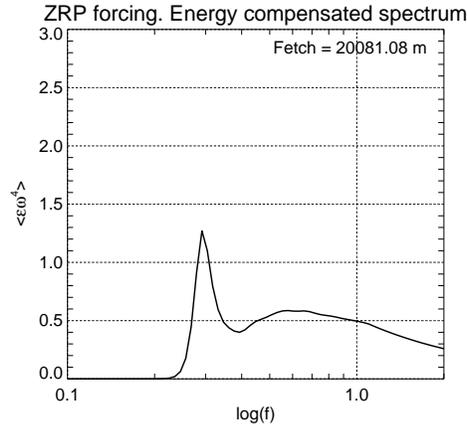}
	\caption{ Angle averaged, energy compensated, spectrum $<\varepsilon \omega^{4}> = \frac{1}{2\pi} \int \varepsilon (\omega,\theta) \omega^{4} d \theta$ as the function of decimal logarithm of frequency $f=\frac{\omega}{2\pi}$ shows the behavior close to the theoretically predicted value $\beta \simeq C_K \left( g^4 P\right)^{1/3}\simeq 0.5$ in the area of the ``plateau".}
	\label{EnComSpec}
\end{figure}

\noindent For the sake of brevity, the calculation of density flux to high wavenumbers is omitted, and only the final result is presented:  $P \simeq 2 \cdot 10^{-6} \,\, m^2/sec$, which gives the value $\beta_{KZ} \simeq 0.5 \,\, m^2/sec^3$. An approximation of $\beta_{KZ}$ is given by angle averaged compensated spectrum $\varepsilon \omega^{-4}$, shown in Fig.\ref{EnComSpec}.

%%%%%%%%%%%%%%%%%%%%%%%%%%%%%%%%%%%%%%%%%%%%%%%%%%%%%%%%%%%%%%%%%%%%%%%%%%%%%%%%%%%%%%%%%%%%%%%%%%

%%%%%%%%%%%%%%%%%%%%%%%%%%%%%%%%%%%%%%%%%%%%%%%%%%%%%%%%%%%%%%%%%%%%%%%%%%%%%%%%%%%%%%%%%%%%%%%%%%

\begin{figure}
        \centering
	\includegraphics[width=0.5\textwidth]{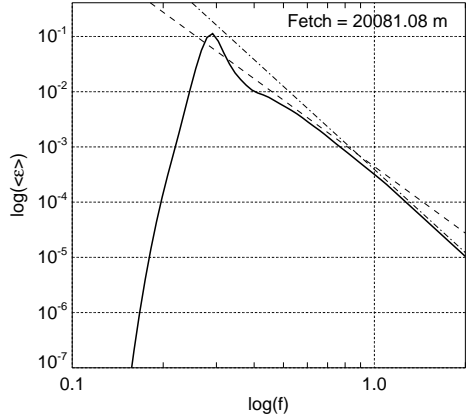}
	\caption{Decimal logarithm of angle averaged spectral energy density $<\varepsilon> = \frac{1}{2\pi} \int_{0}^{2\pi} \epsilon d\theta$, as a function of the decimal logarithm of the frequency $f=\frac{\omega }{2\pi}$ - solid line. Dashed line - fit $\sim \omega ^{-4}$, dash-dotted line - fit $\sim \omega ^{-5}$.}
	\label{SpecA}
\end{figure}

\noindent Finally, Fig.\ref{SpecA} presents angle averaged energy spectrum, as the function of frequency, in logarithmic coordinates. One can see that it consists of segments of:

\begin{enumerate}
\item  Spectral maximum area
\item  Kolmogorov-Zakharov  spectrum $\omega^{-4}$
\item  Phillips high frequency tail $\omega^{-5}$
\end{enumerate}

%%%%%%%%%%%%%%%%%%%%%%%%%%%%%%%%%%%%%%%%%%%%%%%%%%%%%%%%%%%%%%%%%%%%%%%%%%%%%%%%%%%%%%%%%%%%%%%%%%

%%%%%%%%%%%%%%%%%%%%%%%%%%%%%%%%%%%%%%%%%%%%%%%%%%%%%%%%%%%%%%%%%%%%%%%%%%%%%%%%%%%%%%%%%%%%%%%%%%

\noindent It follows from Fig.\ref{EnComSpec}, \ref{SpecA} that, in the interval $0.4 \,\, Hz< f < 1 \,\, Hz$, the energy spectrum is close to the Zakharov-Filonenko spectrum Eq.(\ref{KZsolution}), with the accuracy $20\%$. One cannot expect higher accuracy, due to the anisotropy of the realized spectrum and the influence of the high frequency dissipation, as well. One can state, nevertheless, that in the domain of frequencies exceeding spectral maximum frequencies, the energy spectrum is fairly close to the one described by equation $S_{nl} = 0$, which confirms the view that the energy balance of wind excited surface waves, as presented in current research.

\noindent The analysis carried out in the previous section shows that the quality of the different versions of wind input terms $S_{in}$ should be estimated by the following criteria:

\begin{enumerate}
\item  Checking powers of the observed energy and mean frequency dependencies Eq.(\ref{Peq}), (\ref{Qeq}) along the fetch, versus what is predicted by self-similar solutions.

\item  Checking the ``magic relations'' Eq.(\ref{MagicRelation}) between exponents $p$ and $q$, for observed energy and frequency dependencies, along the fetch.

\item  Checking the exponents of directional (angle averaged) spectral energy dependencies versus the Kolmogorov-Zakharov exponent $-4$.
\end{enumerate}

\noindent We applied such tests to the results of $HE$ simulations, which used the following popular wind input terms, within alternative framework:

\noindent 

\begin{enumerate}

\item  Chalikov $S_{in}$ term \cite{R25,R15}

\item  Snyder $S_{in}$ term \cite{R30}

\item   Hsiao-Shemdin $S_{in}$ term \cite{R31}

\item  $WAM3$ $S_{in}$ term \cite{R52}

\end{enumerate}

\section{Test of Chalikov wind input term} \label{ChalikovTest}

\noindent The sophisticated Chalikov wind input term algorithm is not presented in the current paper, due to space. Curious readers can find it in \cite{R15, R25}.

\begin{figure}
        \centering
        \begin{subfigure}[b]{0.45\textwidth}
                \includegraphics[width=\textwidth]{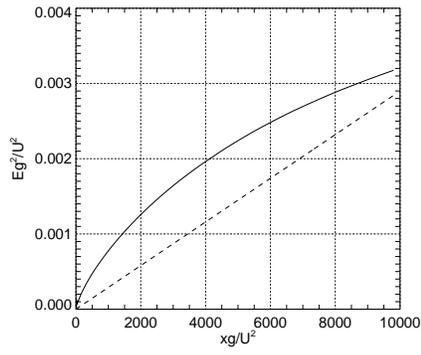}
                \caption{}
                \label{ChalikovFormA}
        \end{subfigure}
\qquad
%(or a blank line to force the sub-figure onto a new line)
        \begin{subfigure}[b]{0.45\textwidth}
                \includegraphics[width=\textwidth]{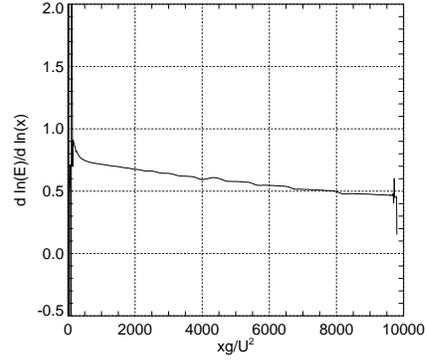}
                \caption{}
                \label{ChalikovFormB}
        \end{subfigure}
	\caption{Same as Fig.\ref{AltForm}, but for Chalikov $S_{in}$}
	\label{ChalikovForm}
\end{figure}

\noindent Fig.\ref{ChalikovFormA} shows that total energy growth, along the fetch, significantly exceeds what is observed in $ZRP$ simulation.  This dependence is not the power function of the fetch, see Fig.\ref{ChalikovFormB}, but can be approximated by relatively slowly changing values of the exponent $p$, ranging from $0.8$ to $0.5$,  along the fetch. While $ p \simeq 0.8$ is still observed in some experiments, the value $p \simeq 0.5$ is completely unrealistic. 

%%%%%%%%%%%%%%%%%%%%%%%%%%%%%%%%%%%%%%%%%%%%%%%%%%%%%%%%%%%%%%%%%%%%%%%%%%%%%%%%%%%%%%%%%%%%%%%%%%

%%%%%%%%%%%%%%%%%%%%%%%%%%%%%%%%%%%%%%%%%%%%%%%%%%%%%%%%%%%%%%%%%%%%%%%%%%%%%%%%%%%%%%%%%%%%%%%%%%

\begin{figure}
        \centering
        \begin{subfigure}[b]{0.45\textwidth}
                \includegraphics[width=\textwidth]{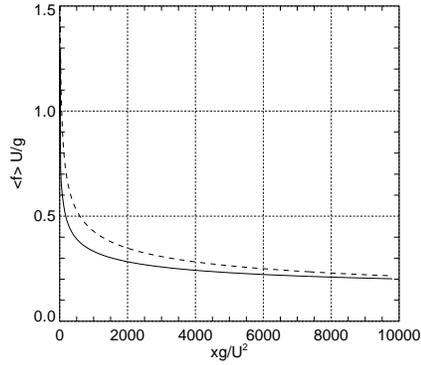}
                \caption{}
                \label{MeanFreqChalikovA}
        \end{subfigure}
\qquad
%(or a blank line to force the sub-figure onto a new line)
        \begin{subfigure}[b]{0.45\textwidth}
                \includegraphics[width=\textwidth]{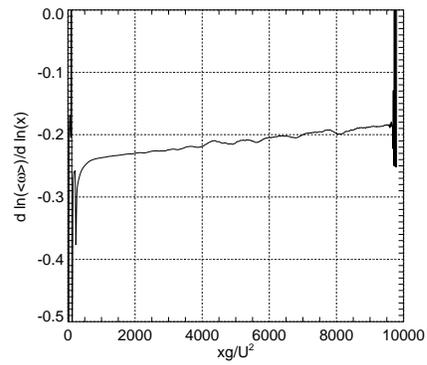}
                \caption{}
                \label{MeanFreqChalikovB}
        \end{subfigure}
	\caption{ Same as Fig.\ref{MeanFreq}, but for Chalikov $S_{in}$ }
	\label{MeanFreqChalikov}
\end{figure}

\noindent The same relates to the mean frequency dependence, against the fetch, shown in Fig.\ref{MeanFreqChalikovA}, with the values of the exponent $q$ shown in Fig.\ref{MeanFreqChalikovB}. The value of $q$ is also not constant, but slowly diminishes with the fetch. One should note that the value $q \simeq 0.25$ has been detected in the experiments, while $q<0.2$ has never, apparently, occurred.

%%%%%%%%%%%%%%%%%%%%%%%%%%%%%%%%%%%%%%%%%%%%%%%%%%%%%%%%%%%%%%%%%%%%%%%%%%%%%%%%%%%%%%%%%%%%%%%%%%

%%%%%%%%%%%%%%%%%%%%%%%%%%%%%%%%%%%%%%%%%%%%%%%%%%%%%%%%%%%%%%%%%%%%%%%%%%%%%%%%%%%%%%%%%%%%%%%%%%

\begin{figure}
        \centering
	\includegraphics[width=0.5\textwidth]{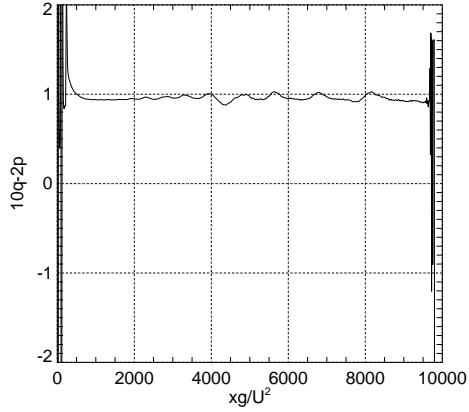}
	\caption{``Magic relation" $(10q-2p)$ as a function of the dimensionless  fetch for Chalikov wind input term.}
	\label{MagicNumberChalikov}
\end{figure}

\noindent Fig.\ref{MagicNumberChalikov} presents the combination, $(10q-2p)$, as a function of the fetch. It is surprising that it is in good accordance with the relation Eq.(\ref{MagicRelation}). It means that despite incorrect values $p$ and $q$ along the fetch, their combination $(10q-2p)$ still holds in complete accordance with theoretical prediction, and the spectra are ``locally self-similar", in accordance with $WTT$.

%%%%%%%%%%%%%%%%%%%%%%%%%%%%%%%%%%%%%%%%%%%%%%%%%%%%%%%%%%%%%%%%%%%%%%%%%%%%%%%%%%%%%%%%%%%%%%%%%%

%%%%%%%%%%%%%%%%%%%%%%%%%%%%%%%%%%%%%%%%%%%%%%%%%%%%%%%%%%%%%%%%%%%%%%%%%%%%%%%%%%%%%%%%%%%%%%%%%%

\begin{figure}
        \centering
	\includegraphics[width=0.5\textwidth]{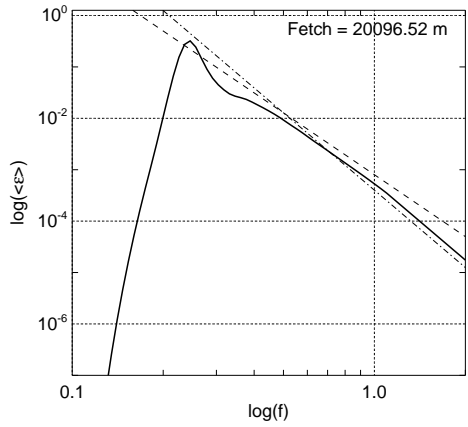}
	\caption{Same as Fig.\ref{SpecA}, but for Chalikov $S_{in}$.}
	\label{SpecChalikovA}
\end{figure}

\noindent Fig.\ref{SpecChalikovA} presents directional (angle averaged) spectrum, as a function of frequency, in logarithmic coordinates. One can see Kolmogorov-Zakharov $\sim \omega ^{-4}$ and Phillips high frequency tail $\sim \omega^{-5}$, as well. 

%%%%%%%%%%%%%%%%%%%%%%%%%%%%%%%%%%%%%%%%%%%%%%%%%%%%%%%%%%%%%%%%%%%%%%%%%%%%%%%%%%%%%%%%%%%%%%%%%%

%%%%%%%%%%%%%%%%%%%%%%%%%%%%%%%%%%%%%%%%%%%%%%%%%%%%%%%%%%%%%%%%%%%%%%%%%%%%%%%%%%%%%%%%%%%%%%%%%%

\begin{figure}
        \centering
	\includegraphics[width=0.5\textwidth]{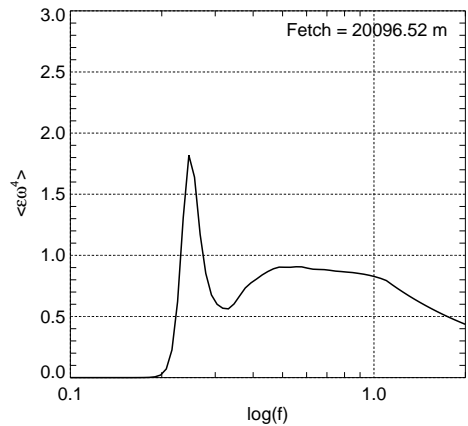}
	\caption{Angle averaged energy compensated spectrum $<\varepsilon \omega^{4}> = \frac{1}{2\pi} \int \varepsilon (\omega,\theta) \omega^{4} d \theta$ as the function of decimal logarithm of frequency $f=\frac{\omega}{2\pi}$  for Chalikov $S_{in}$.}
	\label{Beta_KZ_Chalikov}
\end{figure}

\noindent Fig.\ref{Beta_KZ_Chalikov} confirms the presence of $~\omega^{-4}$ spectrum, through existence of the ``plateau" section, to the right of the spectral peak area, in the frequency range $0.45 Hz < f < 1 Hz$.

%%%%%%%%%%%%%%%%%%%%%%%%%%%%%%%%%%%%%%%%%%%%%%%%%%%%%%%%%%%%%%%%%%%%%%%%%%%%%%%%%%%%%%%%%%%%%%%%%%

%%%%%%%%%%%%%%%%%%%%%%%%%%%%%%%%%%%%%%%%%%%%%%%%%%%%%%%%%%%%%%%%%%%%%%%%%%%%%%%%%%%%%%%%%%%%%%%%%%

\section{Test of Snyder wind input term} \label{SnyderTest}

\noindent Snyder wind input term Eq.(\ref{BetaSHB}) is especially important, since it is included, as an option, in operational models.  

\begin{figure}
        \centering
        \begin{subfigure}[b]{0.45\textwidth}
                \includegraphics[width=\textwidth]{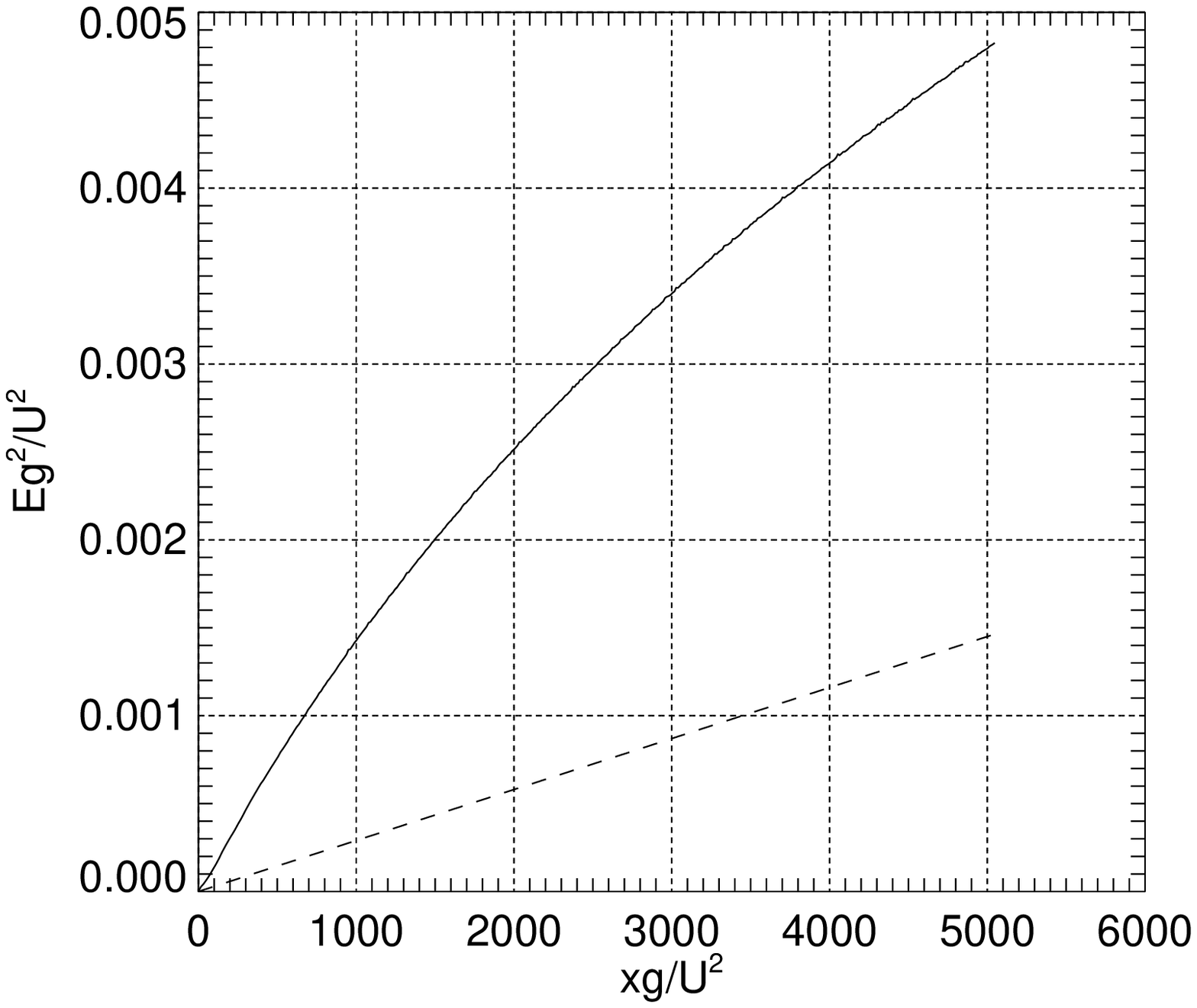}
                \caption{}
                \label{SnyderFormA}
        \end{subfigure}
\qquad
%(or a blank line to force the sub-figure onto a new line)
        \begin{subfigure}[b]{0.45\textwidth}
                \includegraphics[width=\textwidth]{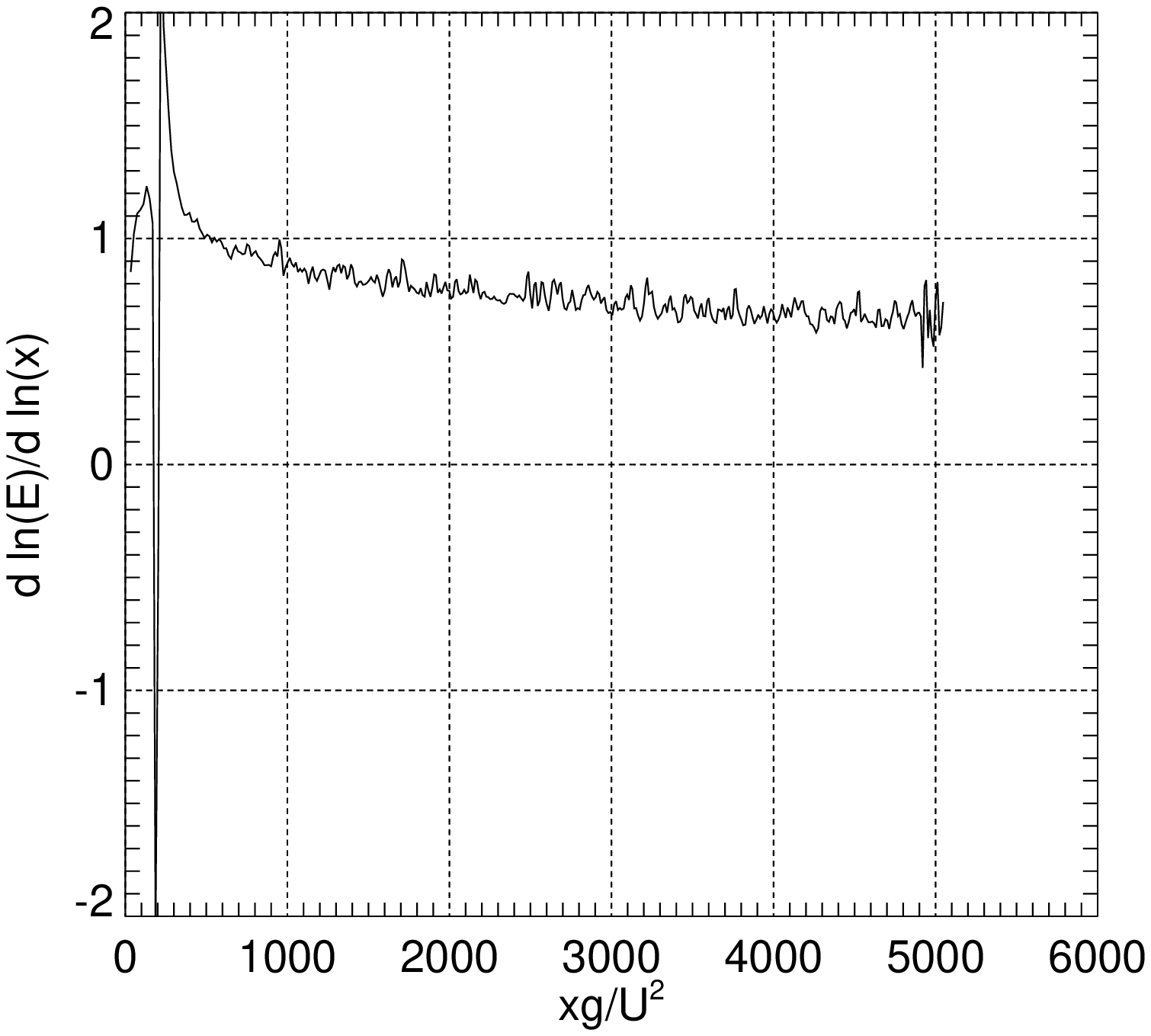}
                \caption{}
                \label{SnyderFormB}
        \end{subfigure}
	\caption{Same as Fig.\ref{AltForm}, but for Snyder $S_{in}$}
	\label{SnyderForm}
\end{figure}

\noindent The main disadvantage of the Snyder wind input term is rapid energy growth, with the fetch, presented in Fig.\ref{SnyderFormA}. For dimensionless fetches  $\chi \simeq 5 \cdot 10^{3}$ it shows values approximately three times bigger than those experimentally observed and those obtained in the more realistic $ZRP$ model. Apparently, that fact, together with a non-critical belief in the Snyder wind input function, caused the myth about long-wave dissipation, due to breaking of the long waves.

\noindent Despite unrealistic energy growth along the fetch, Fig.\ref{SnyderFormB} shows that energy growth is close to the power function Eq.(\ref{Peq}), with the index $p$ slowly changing from $1$ to $0.6$, which is, generally, significantly lower than observed in field experiments.

%%%%%%%%%%%%%%%%%%%%%%%%%%%%%%%%%%%%%%%%%%%%%%%%%%%%%%%%%%%%%%%%%%%%%%%%%%%%%%%%%%%%%%%%%%%%%%%%%%

%%%%%%%%%%%%%%%%%%%%%%%%%%%%%%%%%%%%%%%%%%%%%%%%%%%%%%%%%%%%%%%%%%%%%%%%%%%%%%%%%%%%%%%%%%%%%%%%%%

\begin{figure}
        \centering
        \begin{subfigure}[b]{0.45\textwidth}
                \includegraphics[width=\textwidth]{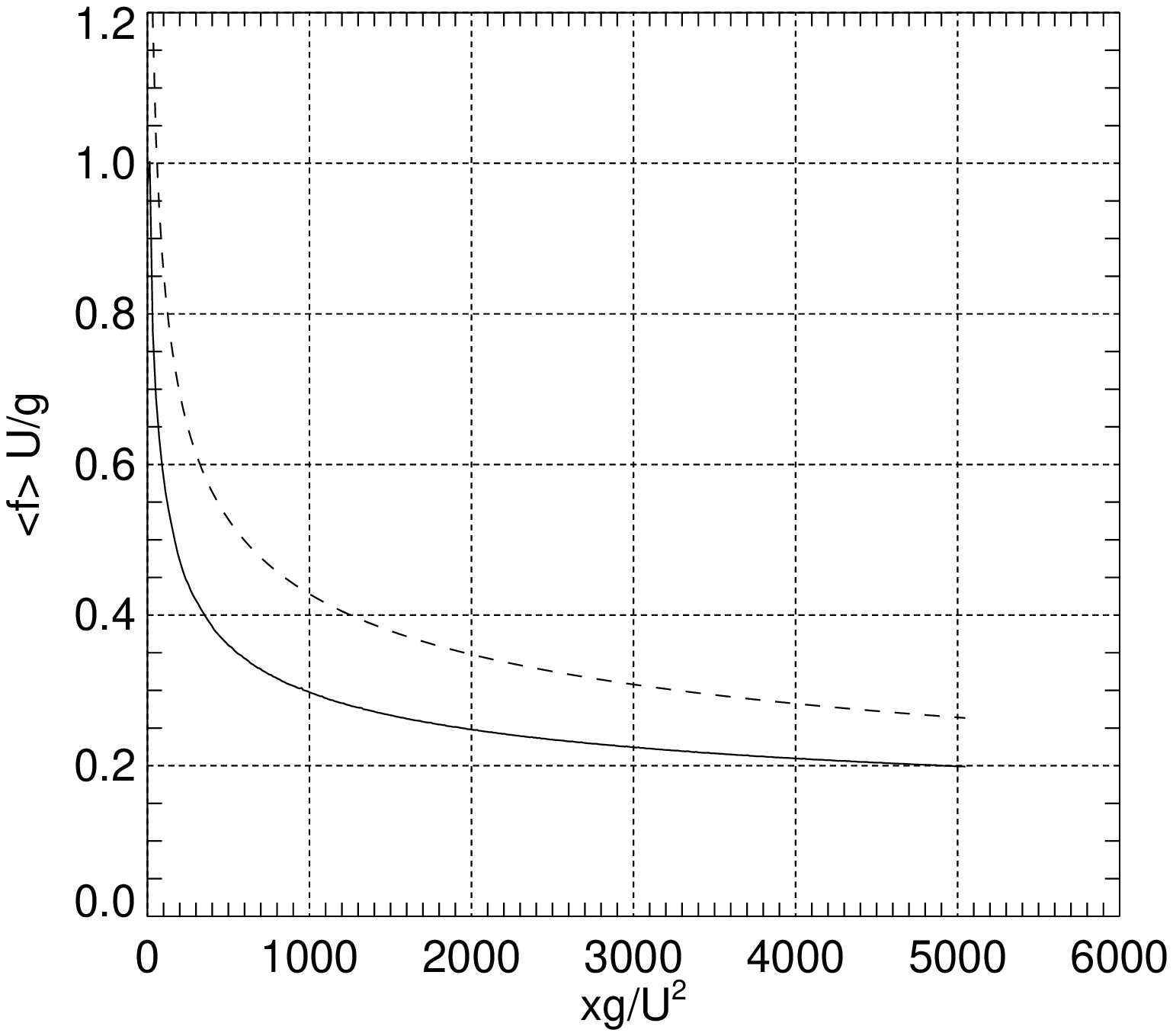}
                \caption{}
                \label{MeanFreqSnyderA}
        \end{subfigure}
\qquad
%(or a blank line to force the sub-figure onto a new line)
        \begin{subfigure}[b]{0.45\textwidth}
                \includegraphics[width=\textwidth]{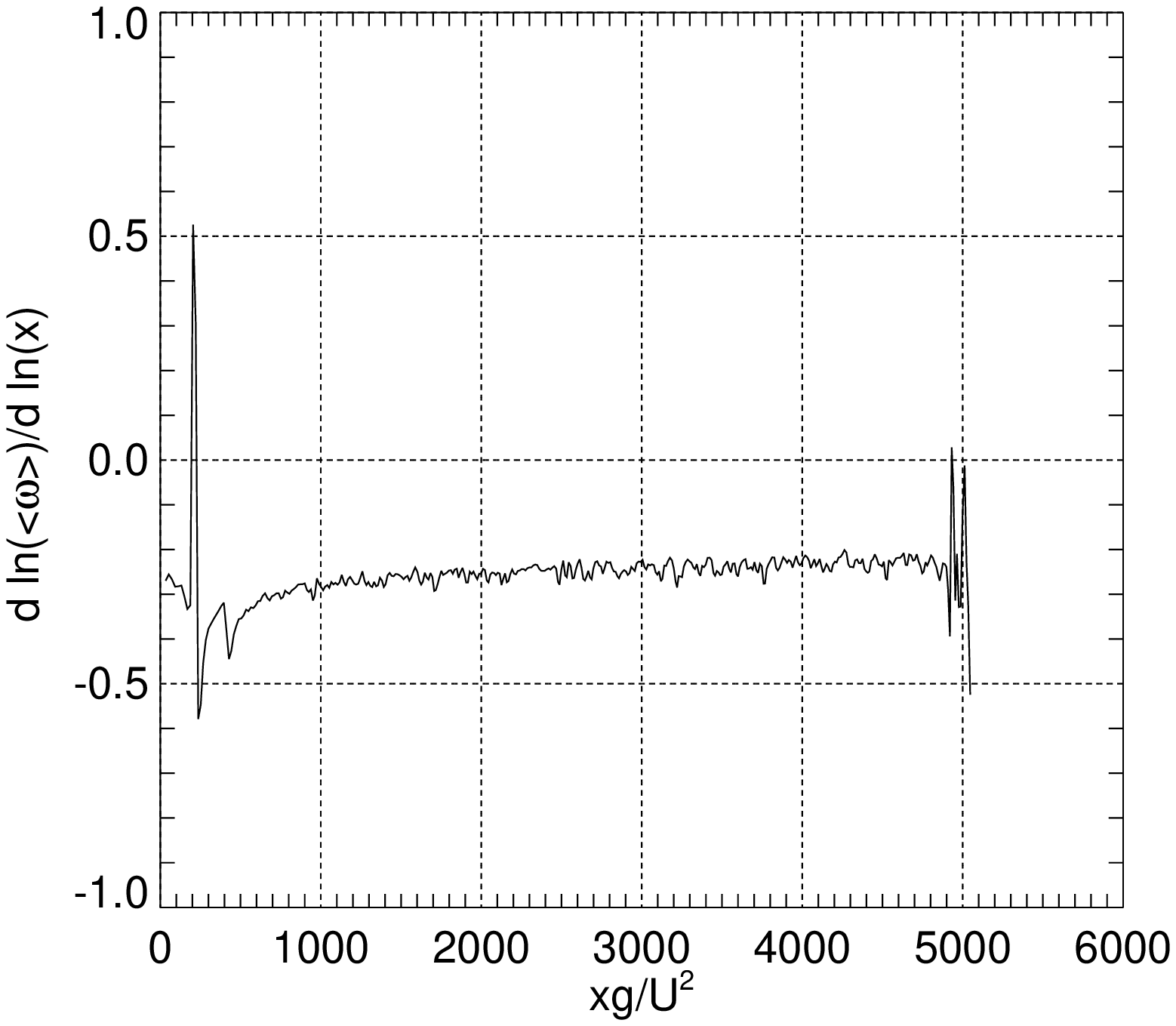}
                \caption{}
                \label{MeanFreqSnyderB}
        \end{subfigure}
      \caption{Same as Fig.\ref{MeanFreq}, but for Snyder $S_{in}$}\label{MeanFreqSnyder}
\end{figure}

\noindent The same relates to the mean frequency dependence, against the fetch, shown in Fig.\ref{MeanFreqSnyderA}, with the values of the exponent $q$ shown in Fig.\ref{MeanFreqSnyderB}. The value of $q$ is not constant, but slowly diminishes with the fetch. One should note that the value $q \simeq 0.25$ has been detected in experiments, while $q<0.2$ has never, apparently, occurred.

\noindent Dependence of the mean frequency, against the fetch, shown in Fig.\ref{MeanFreqSnyderA}, is lower than $ZRP$ numerical results, but can be also approximated by a power function of the fetch Eq.(\ref{Qeq}), with the values of $q$ slowly diminishing along the fetch, from $0.3$ to $0.25$, see Fig.\ref{MeanFreqSnyderB}.

%%%%%%%%%%%%%%%%%%%%%%%%%%%%%%%%%%%%%%%%%%%%%%%%%%%%%%%%%%%%%%%%%%%%%%%%%%%%%%%%%%%%%%%%%%%%%%%%%%

%%%%%%%%%%%%%%%%%%%%%%%%%%%%%%%%%%%%%%%%%%%%%%%%%%%%%%%%%%%%%%%%%%%%%%%%%%%%%%%%%%%%%%%%%%%%%%%%%%

\begin{figure}
        \centering
	\includegraphics[width=0.5\textwidth]{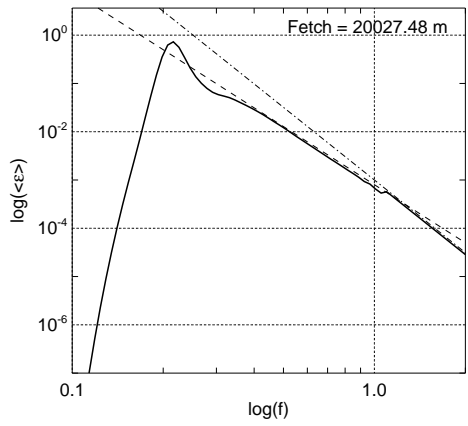}
	\caption{Same as Fig.\ref{SpecA}, but for Snyder $S_{in}$.}
	\label{SpecSnyder}
\end{figure}

\begin{figure}
        \centering
	\includegraphics[width=0.5\textwidth]{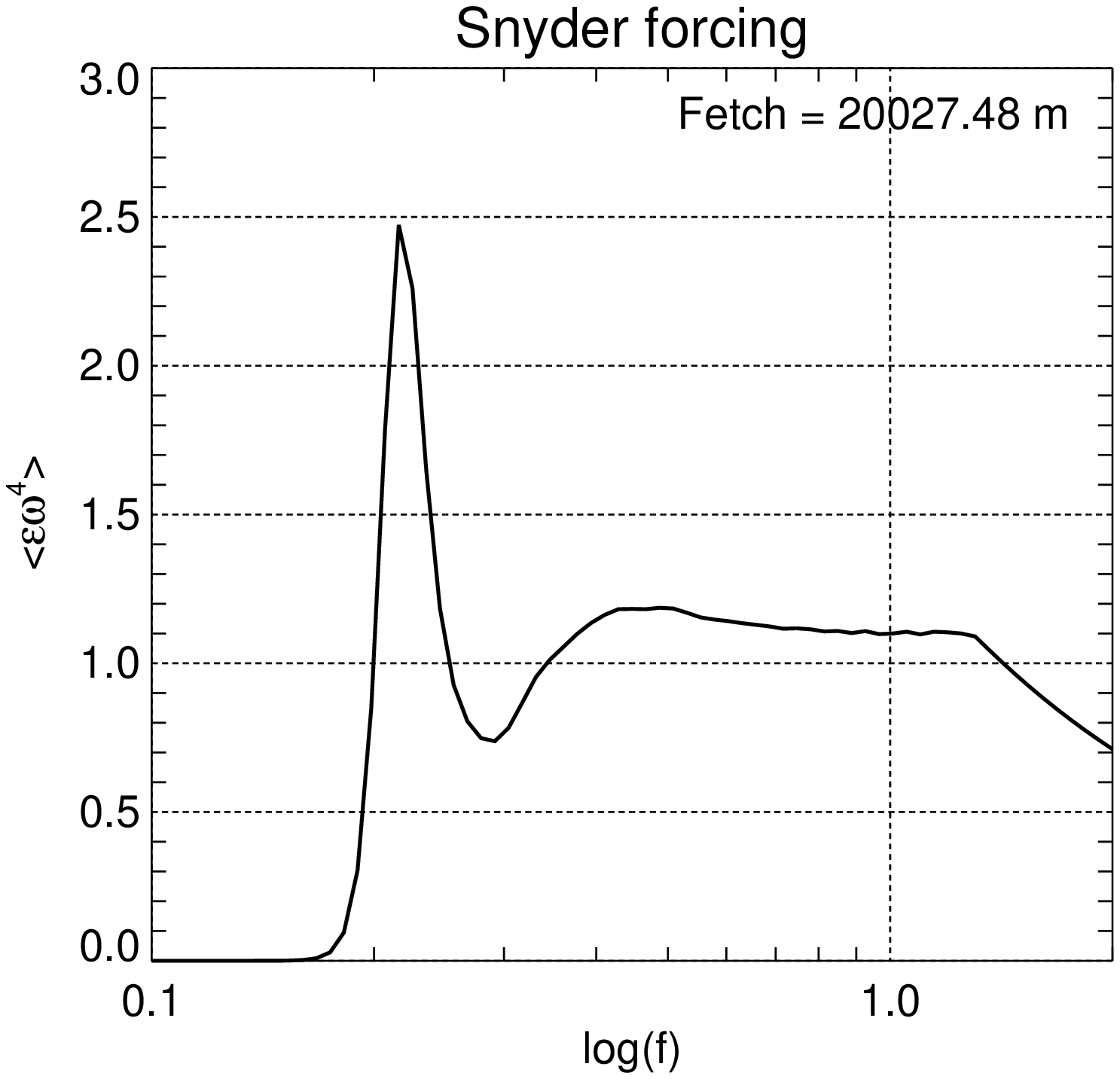}
	\caption{Angle averaged, energy compensated, spectrum $<\varepsilon \omega^{4}> = \frac{1}{2\pi} \int \varepsilon (\omega,\theta) \omega^{4} d \theta$ as the function of decimal logarithm of the frequency $f=\frac{\omega}{2\pi}$,  for Snyder $S_{in}$.}
	\label{BetaKZ_Snyder}
\end{figure}

\noindent The Kolmogorov-Zakharov spectrum $\sim \omega ^{-4}$ and Phillips high frequency tail $\sim \omega ^{-5}$ can be seen in Fig.\ref{SpecSnyder}, presenting directional spectrum as a function of frequency, in logarithmic coordinates. The span of the Kolmogorov-Zakharov $\sim \omega ^{-4}$ segment can be estimated from  Fig.\ref{BetaKZ_Snyder}.

%%%%%%%%%%%%%%%%%%%%%%%%%%%%%%%%%%%%%%%%%%%%%%%%%%%%%%%%%%%%%%%%%%%%%%%%%%%%%%%%%%%%%%%%%%%%%%%%%%

%%%%%%%%%%%%%%%%%%%%%%%%%%%%%%%%%%%%%%%%%%%%%%%%%%%%%%%%%%%%%%%%%%%%%%%%%%%%%%%%%%%%%%%%%%%%%%%%%%

%%%%%%%%%%%%%%%%%%%%%%%%%%%%%%%%%%%%%%%%%%%%%%%%%%%%%%%%%%%%%%%%%%%%%%%%%%%%%%%%%%%%%%%%%%%%%%%%%%

%%%%%%%%%%%%%%%%%%%%%%%%%%%%%%%%%%%%%%%%%%%%%%%%%%%%%%%%%%%%%%%%%%%%%%%%%%%%%%%%%%%%%%%%%%%%%%%%%%

\begin{figure}
        \centering
	\includegraphics[width=0.5\textwidth]{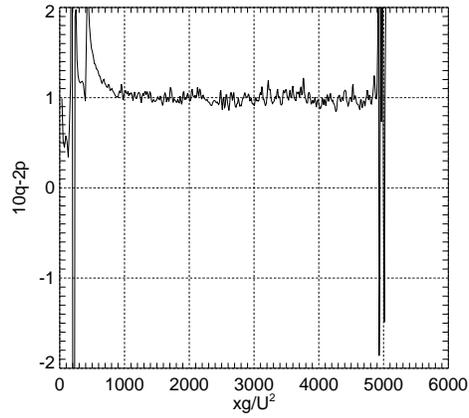}
	\caption{Relation $(10q-2p)$ as a function of the fetch $x$ for Snyder wind input term.}
	\label{MagicNumberSnyder}
\end{figure}

\noindent Fig.\ref{MagicNumberSnyder} presents the combination $(10q-2p)$ as a function of the fetch. Again, it strongly agrees with the theoretical relation Eq.(\ref{MagicRelation}). As in the Chalikov case, it means that, despite imperfect values of $p$ and $q$ and rapid energy growth along the fetch, their combination $(10q-2p)$ still holds in complete accordance with theoretical prediction, i.e. self-similarity is also fulfilled, locally, in the Snyder case.

%%%%%%%%%%%%%%%%%%%%%%%%%%%%%%%%%%%%%%%%%%%%%%%%%%%%%%%%%%%%%%%%%%%%%%%%%%%%%%%%%%%%%%%%%%%%%%%%%%

%%%%%%%%%%%%%%%%%%%%%%%%%%%%%%%%%%%%%%%%%%%%%%%%%%%%%%%%%%%%%%%%%%%%%%%%%%%%%%%%%%%%%%%%%%%%%%%%%%

\section{Test of \textit{Hsiao-Shemdin} wind input term} \label{HStest}

\begin{figure}
        \centering
        \begin{subfigure}[b]{0.45\textwidth}
                \includegraphics[width=\textwidth]{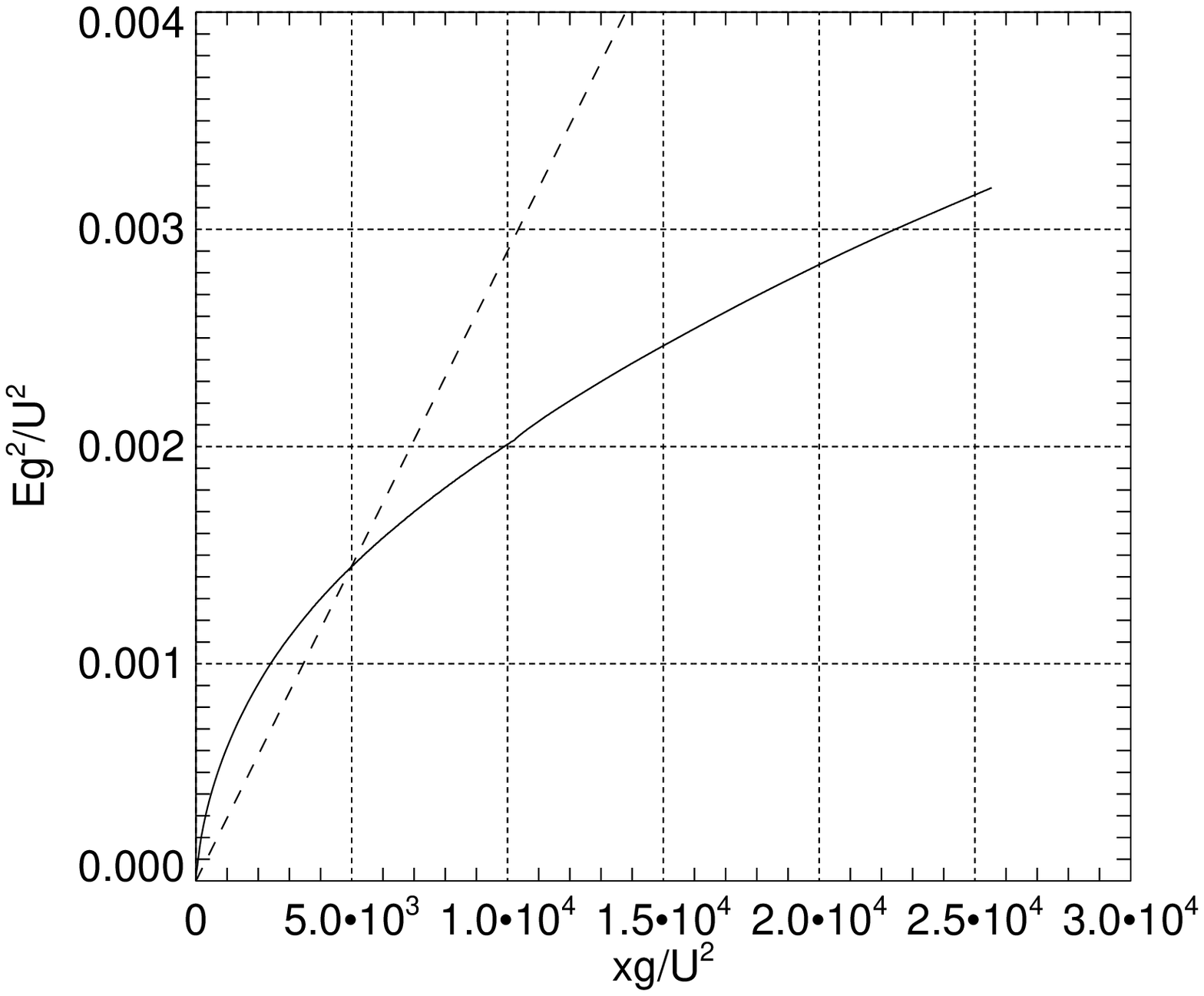}
                \caption{}
                \label{HSFormA}
        \end{subfigure}
\qquad
%(or a blank line to force the sub-figure onto a new line)
        \begin{subfigure}[b]{0.45\textwidth}
                \includegraphics[width=\textwidth]{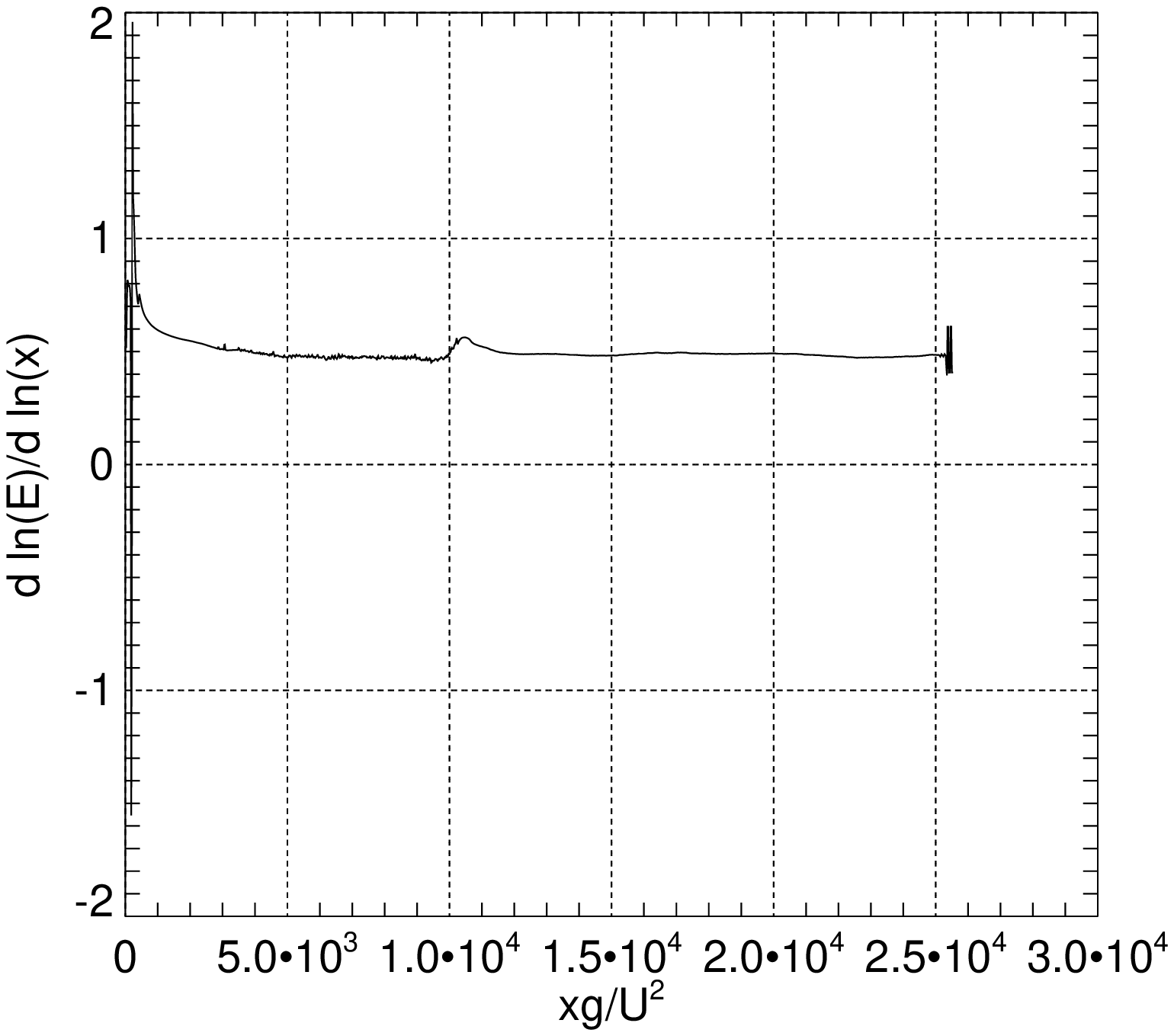}
                \caption{}
                \label{HSFormB}
        \end{subfigure}
	\caption{Same as Fig.\ref{AltForm}, but for Hsiao-Shemdin $S_{in}$}
	\label{HSForm}
\end{figure}

\noindent Fig.\ref{HSFormA} shows that total energy growth, along the fetch, underestimates $ZRP$ simulation. It obeys the power law Eq.(\ref{Peq}) with the exponent $p \approx 0.5$, see Fig.\ref{HSFormB}.

%%%%%%%%%%%%%%%%%%%%%%%%%%%%%%%%%%%%%%%%%%%%%%%%%%%%%%%%%%%%%%%%%%%%%%%%%%%%%%%%%%%%%%%%%%%%%%%%%%

%%%%%%%%%%%%%%%%%%%%%%%%%%%%%%%%%%%%%%%%%%%%%%%%%%%%%%%%%%%%%%%%%%%%%%%%%%%%%%%%%%%%%%%%%%%%%%%%%%

\begin{figure}
        \centering
        \begin{subfigure}[b]{0.45\textwidth}
                \includegraphics[width=\textwidth]{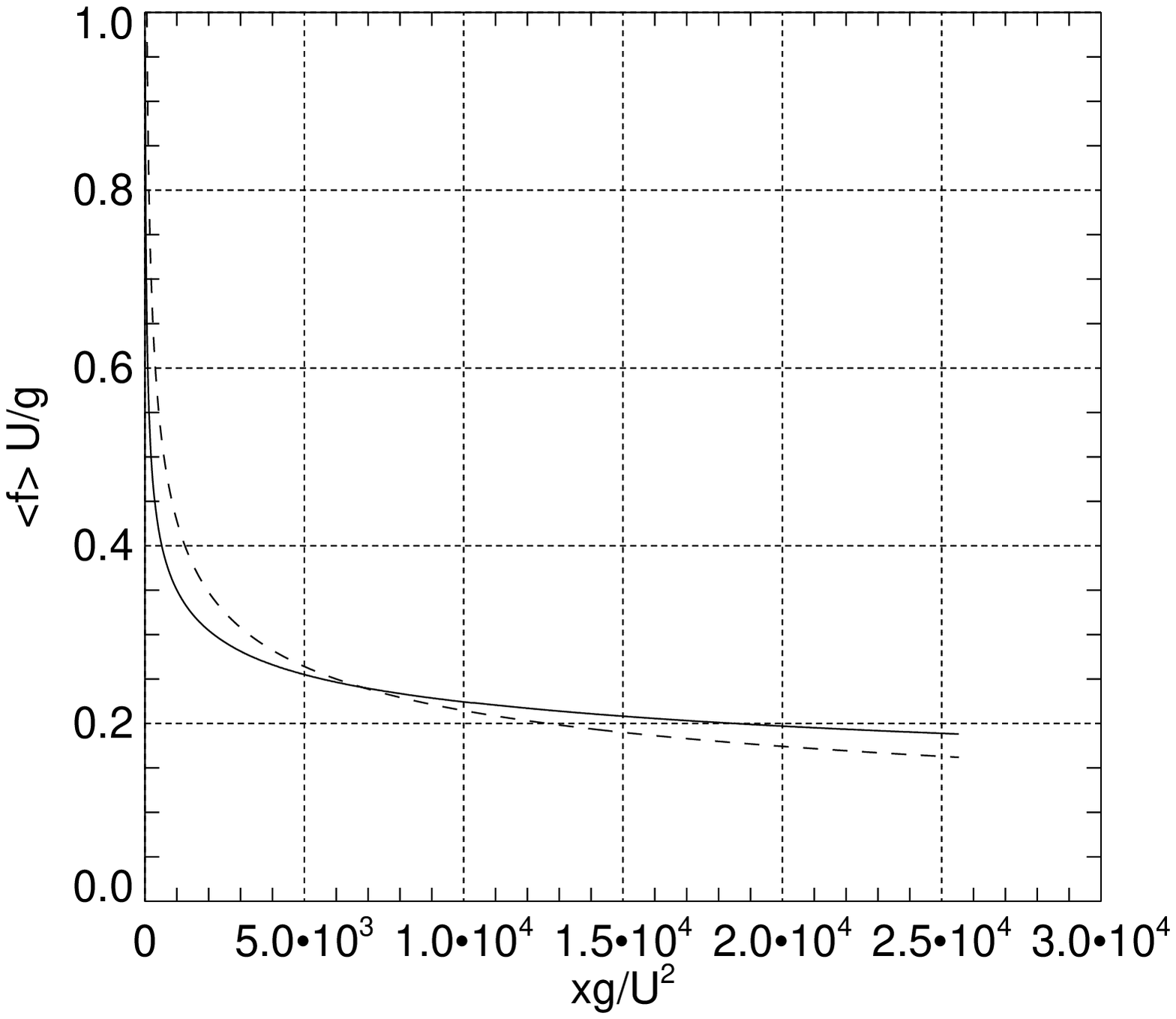}
                \caption{}
                \label{MeanFreqHSA}
        \end{subfigure}
\qquad
%(or a blank line to force the sub-figure onto a new line)
        \begin{subfigure}[b]{0.45\textwidth}
                \includegraphics[width=\textwidth]{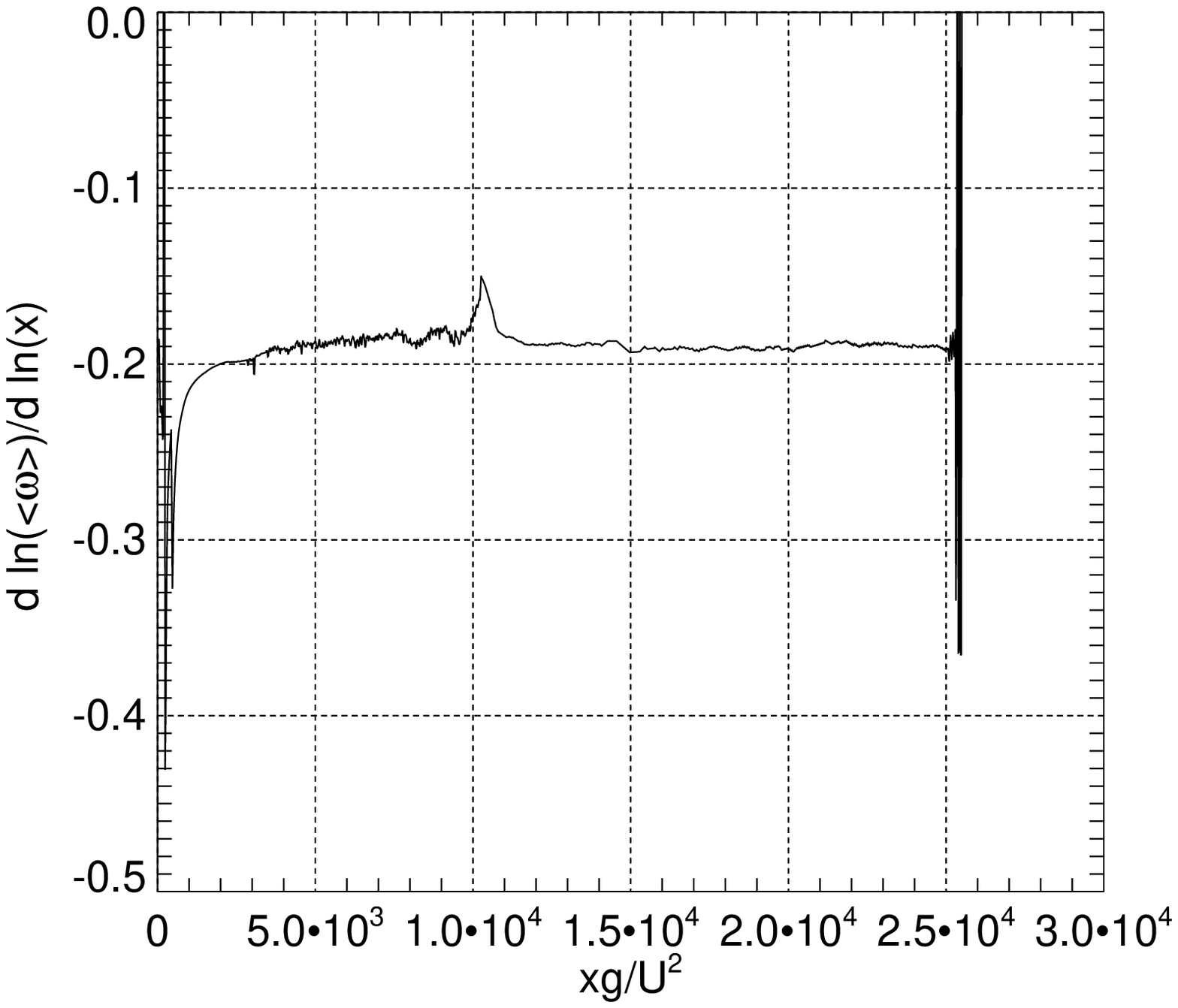}
                \caption{}
                \label{MeanFreqHSB}
	\end{subfigure}
	\caption{Same as Fig.\ref{MeanFreq}, but for Hsiao-Shemdin $S_{in}$}
	\label{MeanFreqHS}
\end{figure}

\noindent  Fig.\ref{MeanFreqHSA} demonstrates mean frequency dependence, on the fetch, by power law Eq.(\ref{Qeq}), with asymptotic value of the index $q \approx 0.21$, see Fig.\ref{MeanFreqHSB}.

%%%%%%%%%%%%%%%%%%%%%%%%%%%%%%%%%%%%%%%%%%%%%%%%%%%%%%%%%%%%%%%%%%%%%%%%%%%%%%%%%%%%%%%%%%%%%%%%%%

%%%%%%%%%%%%%%%%%%%%%%%%%%%%%%%%%%%%%%%%%%%%%%%%%%%%%%%%%%%%%%%%%%%%%%%%%%%%%%%%%%%%%%%%%%%%%%%%%%

\begin{figure}
        \centering
	\includegraphics[width=0.5\textwidth]{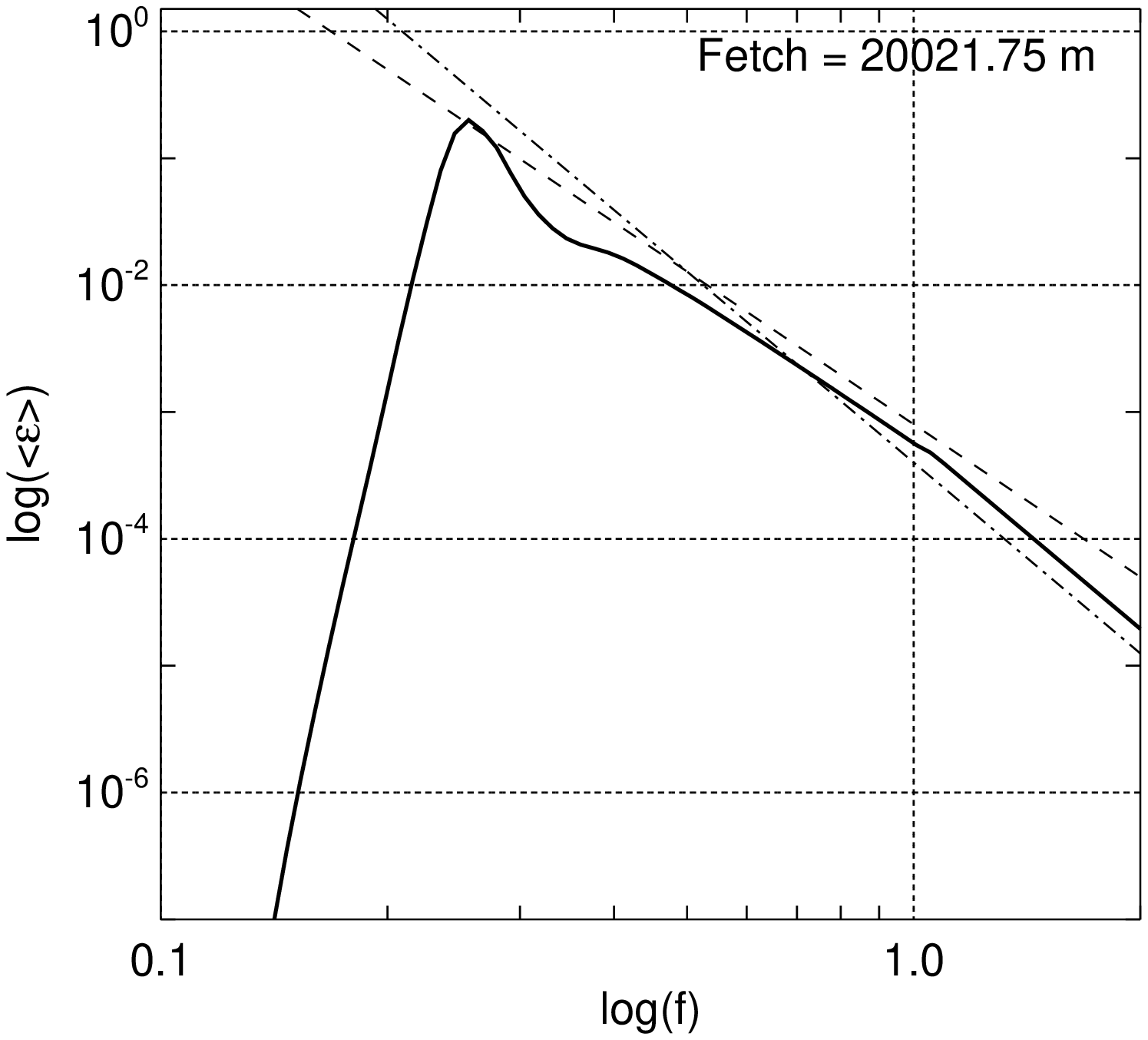}
	\caption{Same as Fig.\ref{SpecA}, but for Hsiao-Shemdin $S_{in}$.}
	\label{SpecHS}
\end{figure}

\begin{figure}
        \centering
	\includegraphics[width=0.5\textwidth]{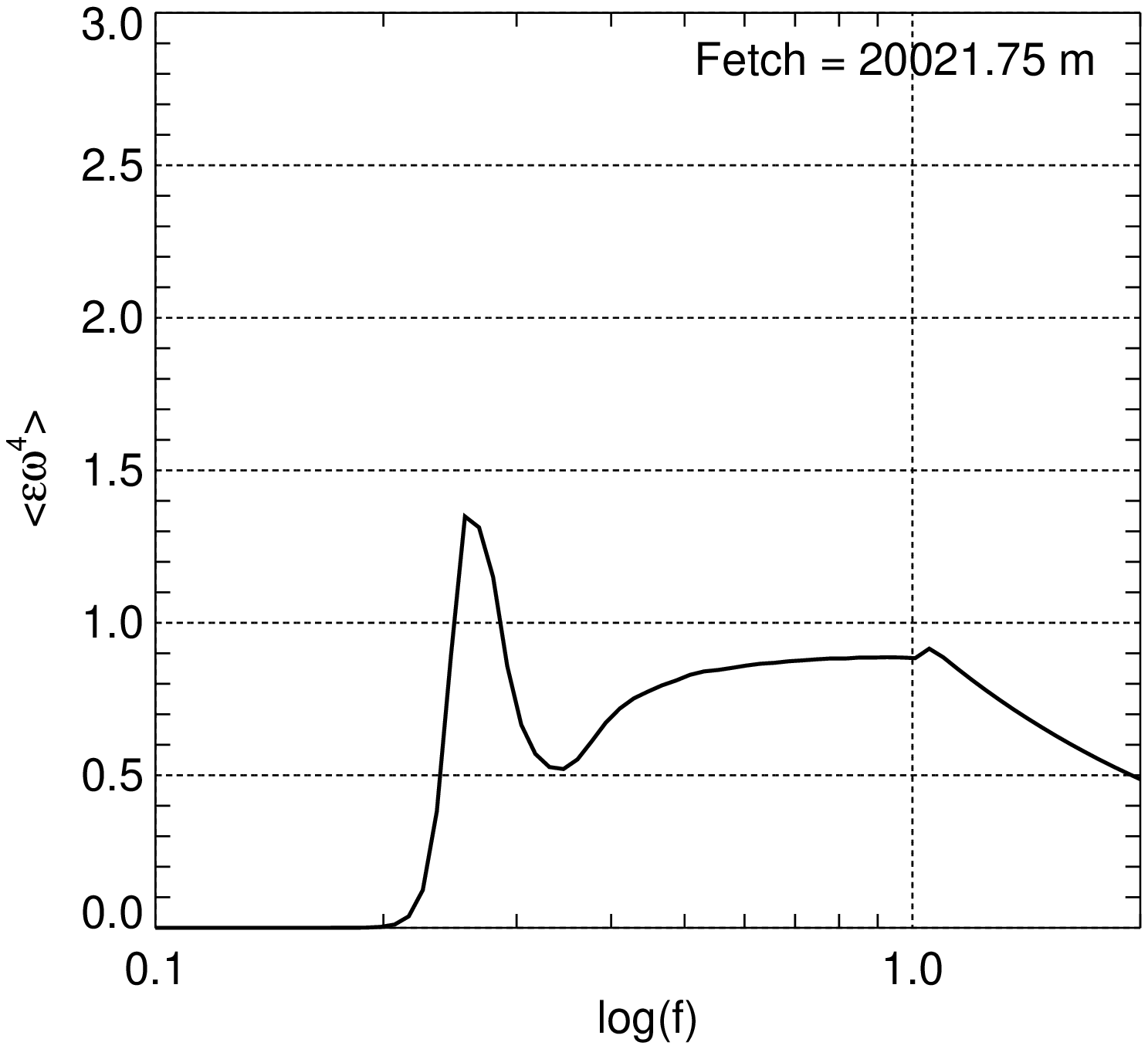}
	\caption{Angle averaged, energy compensated, spectrum $<\varepsilon \omega^{4}> = \frac{1}{2\pi} \int \varepsilon (\omega,\theta) \omega^{4} d \theta$ as the function of decimal logarithm of the frequency $f = \frac{\omega}{2\pi}$  for Hsiao-Shemdin $S_{in}$.}
	\label{BetaKZ_HS}
\end{figure}

\noindent Kolmogorov-Zakharov segment $\sim \omega^{-4}$ and Phillips high frequency tail $\sim \omega^{-5}$ can be seen in Fig.\ref{SpecHS}, presenting directional spectrum as a function of frequency, in logarithmic coordinates. The span of the Kolmogorov-Zakharov spectrum can be estimated using Fig.\ref{BetaKZ_HS}.

%%%%%%%%%%%%%%%%%%%%%%%%%%%%%%%%%%%%%%%%%%%%%%%%%%%%%%%%%%%%%%%%%%%%%%%%%%%%%%%%%%%%%%%%%%%%%%%%%%

%%%%%%%%%%%%%%%%%%%%%%%%%%%%%%%%%%%%%%%%%%%%%%%%%%%%%%%%%%%%%%%%%%%%%%%%%%%%%%%%%%%%%%%%%%%%%%%%%%

%%%%%%%%%%%%%%%%%%%%%%%%%%%%%%%%%%%%%%%%%%%%%%%%%%%%%%%%%%%%%%%%%%%%%%%%%%%%%%%%%%%%%%%%%%%%%%%%%%

\begin{figure}
        \centering
	\includegraphics[width=0.5\textwidth]{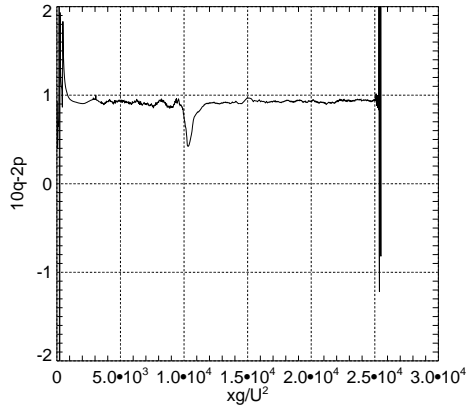}
	\caption{ Relation $(10q-2p)$ as the function of the fetch for Hsiao-Shemdin wind input term.}
	\label{MagicNumberHS}
\end{figure}

\noindent Fig.\ref{MagicNumberHS} presents combination $(10q-2p)$ as the function of the fetch. It is in total agreement with the theoretical predictions Eq.(\ref{MagicRelation}), which means that self-similarity is also fulfilled locally in Hsiao-Shemdin case.

%%%%%%%%%%%%%%%%%%%%%%%%%%%%%%%%%%%%%%%%%%%%%%%%%%%%%%%%%%%%%%%%%%%%%%%%%%%%%%%%%%%%%%%%%%%%%%%%%%

%%%%%%%%%%%%%%%%%%%%%%%%%%%%%%%%%%%%%%%%%%%%%%%%%%%%%%%%%%%%%%%%%%%%%%%%%%%%%%%%%%%%%%%%%%%%%%%%%%

\section{Test of \textit{WAM3} input terms} \label{WAM3test}

\noindent The current section presents what is, arguably, the most important part of the paper. It concerns the numerical simulation performed in the frame of $WAM3$ model \cite{R54,R55,R52}, using exact expression for $S_{nl}$ term. Similar experiments were performed by different authors, more than twenty years ago, see the monograph \cite{R11}. The results presented in the current paper do not contradict them (see Fig.\ref{EnergyWAM3Compared} ) and reveal some new features.

\noindent The source term for WAM cycles 1 through 3 contain, not only wind input term, but also long-wave dissipation \cite{R54,R55,R52}.

\noindent The input source term was used in Snyder form, as per \cite{R52}:

\begin{eqnarray}
S_{in}(k,\theta) = C_{in} \frac{\rho_a}{\rho_w} max\left[0,\left(\frac{28 u_\star}{c}cos(\theta-\theta_w)-1\right)\right]\omega \varepsilon(k,\theta) \\
u_\star=u_{10} \sqrt{(0.8+0.065 u_{10}) 10^{-3}}
\end{eqnarray}
where $C_{in}=0.25$, $\rho_a$ and $\rho_w$ are the densities of air and water, $u_\star$ is the wind friction velocity, and $c$ is the wave phase velocity. 

\noindent White capping dissipation was defined by Eq.(\ref{DissWAM}) \cite{R52}. Turning on such dissipation radically changes the whole physical picture of the dissipation-free Snyder case, both quantitatively and qualitatively as well.

\begin{figure}
        \centering
	\includegraphics[width=0.5\textwidth]{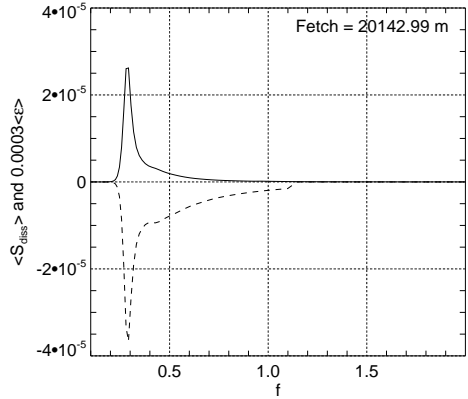}
	\caption{Angle averaged energy dissipation function $<S_{diss}> = \frac{1}{2\pi}\int_0^{2\pi} \gamma_{diss} \varepsilon(\omega,\theta) d\theta$ (dotted line) and angle averaged scaled energy spectrum $<\varepsilon> = \frac{1}{2\pi}\int_0^{2\pi} \varepsilon(\omega,\theta) d\theta)$ (solid line) as the functions of the frequency $f=\frac{1}{2\pi}$.}
	\label{DissVsEnergy}
\end{figure}

\begin{figure}
        \centering
	\includegraphics[width=0.5\textwidth]{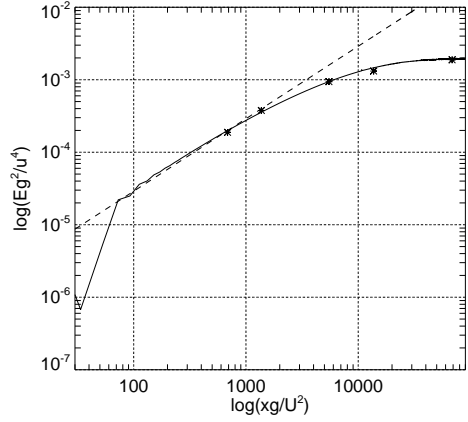}
	\caption{Decimal logarithm of total dimensionless energy, as a function of decimal logarithm, of the dimensional fetch. Solid line -- $WAM3$ case, dashed line -- $ZRP$ case approximated by the fit $2.9 \cdot 10^{-7} \frac{xg}{U^2}$, stars -- data from Fig.3.7 in monograph \cite{R11} recalculated in the assumption $u_* = \frac{1}{27} U_{10}$.}
	\label{EnergyWAM3Compared}
\end{figure}

\noindent First, the dissipation maximum coincides with the spectral maximum. Fig.\ref{DissVsEnergy} demonstrating that $WAM3$ dissipation can be called the ``spectral peak dissipation", indeed due to unambiguous, spectral peak frequency area localization.

%%%%%%%%%%%%%%%%%%%%%%%%%%%%%%%%%%%%%%%%%%%%%%%%%%%%%%%%%%%%%%%%%%%%%%%%%%%%%%%%%%%%%%%%%%%%%%%%%%

%%%%%%%%%%%%%%%%%%%%%%%%%%%%%%%%%%%%%%%%%%%%%%%%%%%%%%%%%%%%%%%%%%%%%%%%%%%%%%%%%%%%%%%%%%%%%%%%%%

\begin{figure}
        \centering
        \begin{subfigure}[b]{0.45\textwidth}
                \includegraphics[width=\textwidth]{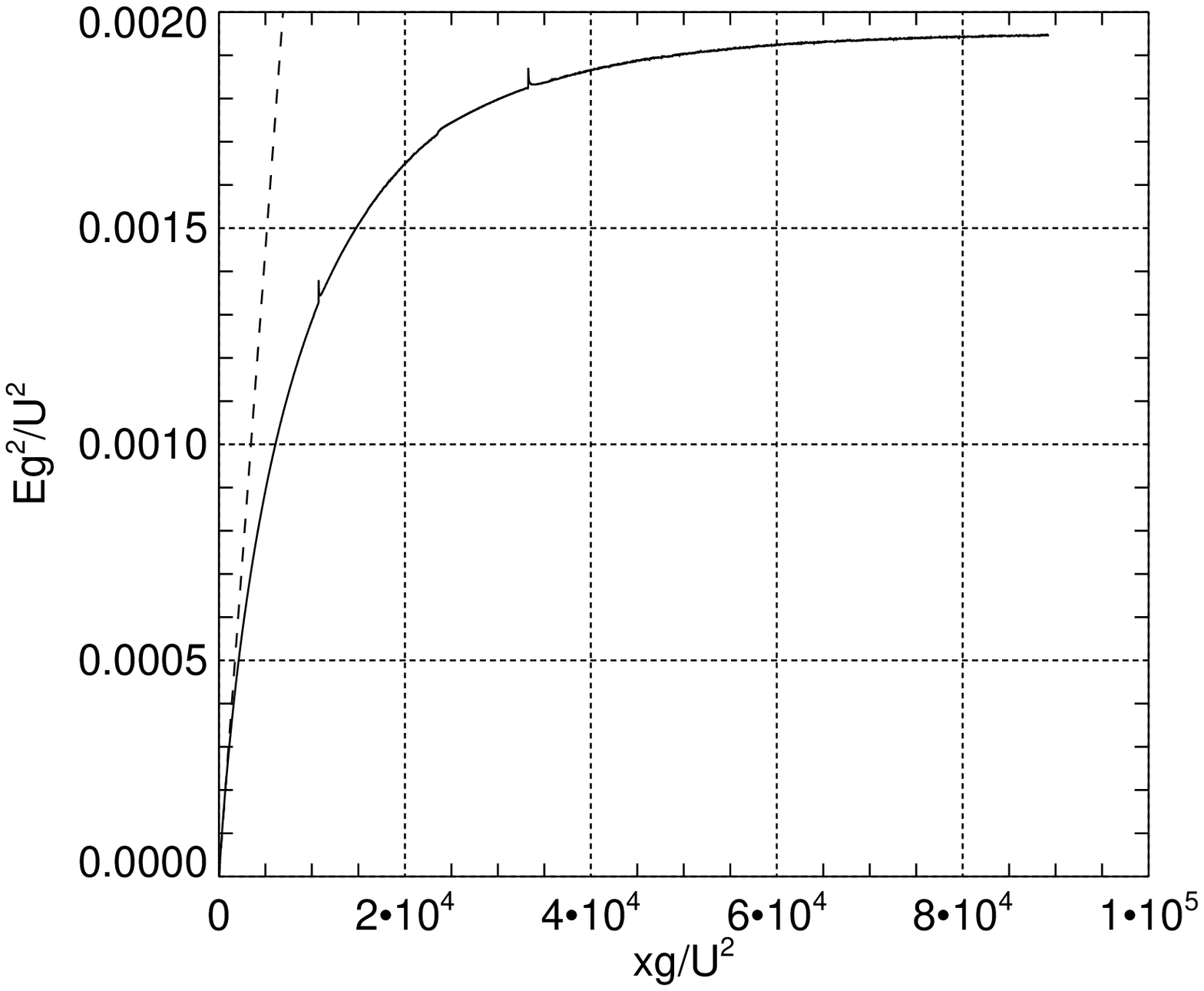}
                \caption{}
                \label{WAM3FormA}
        \end{subfigure}
\qquad
%(or a blank line to force the sub-figure onto a new line)
        \begin{subfigure}[b]{0.45\textwidth}
                \includegraphics[width=\textwidth]{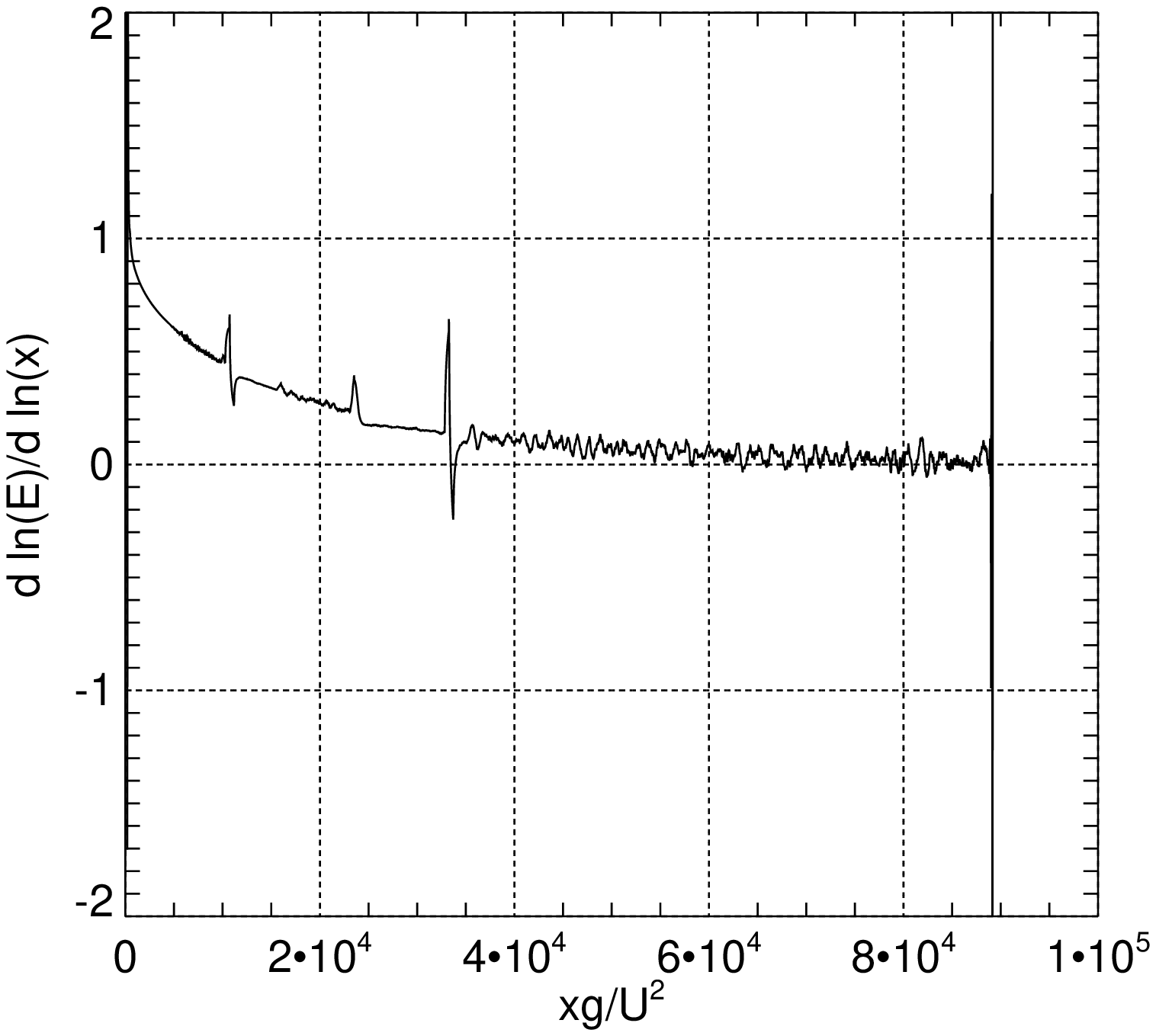}
                \caption{}
                \label{WAM3FormB}
        \end{subfigure}
	\caption{Same as Fig.\ref{AltForm}, but for $WAM3$ $S_{in}$}
	\label{WAM3Form}
\end{figure}

\noindent Fig.\ref{WAM3FormA} shows that total energy growth, along the fetch, becomes constant, at the dimensionless fetch value $\chi=5 \cdot 10^{4}$, which for wind speed $U=10 \,\, m/sec$ means $500 \,\, km$ dimensional fetch. 

\noindent The $WAM3$ model predicts saturation and formation of the ``mature sea". The limiting level of energy $\varepsilon ~ 1.9 \cdot 10^{-3}$ is half that predicted by Pierson-Moscowitz $\varepsilon_{max} \simeq  3.64 \cdot 10^{-3}$. One should notice that Donelan has predicted $\varepsilon_{max} \simeq  4.07 \cdot 10^{-3}$, Young has predicted $\varepsilon_{max} \simeq  (3.6 \pm 0.9 \cdot) 10^{-3}$. Hence, the $WAM3$ model essentially underestimates the energy growth, for large values of the fetch. The exact same results are described in the monograph \cite{R11} (see the corresponding Fig.3.7). 

\noindent The novelty is the following -- for moderate fetches, $10 < \chi < 10^3$, the $WAM3$ model gives the same results as $ZRP$ model, \emph {without any spectral dissipation}. This statement is illustrated by Fig.\ref{EnergyWAM3Compared}, Fig\ref{WAM3FormA}.

%\begin{figure}
%        \centering
%	\includegraphics[width=0.5\textwidth]{fig31.eps}
%	\caption{Decimal logarithm of total dimensionless energy as a function %of decimal logarithm of the dimensional fetch. Solid line -- $WAM3$ case, %dashed line -- $ZRP$ case, stars -- data from Fig.3.7 in monograph \cite{}$ %recalculated in the assumption $u_* = \frac{1}{27} u_{10}$.}
%	\label{EnergyWAM3Compared}
%\end{figure}

\noindent The value of the exponent $p$ versus the fetch, asymptotically, goes to $0$ , see Fig.\ref{WAM3FormB}. This demonstrates strong discrepancies, with $ZRP$ results, for large fetches.

\noindent The stationary level of energy corresponds to $0.2 \,\, m^2$, which is approximately $1.5$ times less than $0.36 \,\, m^2$, the corresponding Pierson-Moscowitz spectrum.

%%%%%%%%%%%%%%%%%%%%%%%%%%%%%%%%%%%%%%%%%%%%%%%%%%%%%%%%%%%%%%%%%%%%%%%%%%%%%%%%%%%%%%%%%%%%%%%%%%

%%%%%%%%%%%%%%%%%%%%%%%%%%%%%%%%%%%%%%%%%%%%%%%%%%%%%%%%%%%%%%%%%%%%%%%%%%%%%%%%%%%%%%%%%%%%%%%%%%

\begin{figure}
        \centering
        \begin{subfigure}[b]{0.45\textwidth}
                \includegraphics[width=\textwidth]{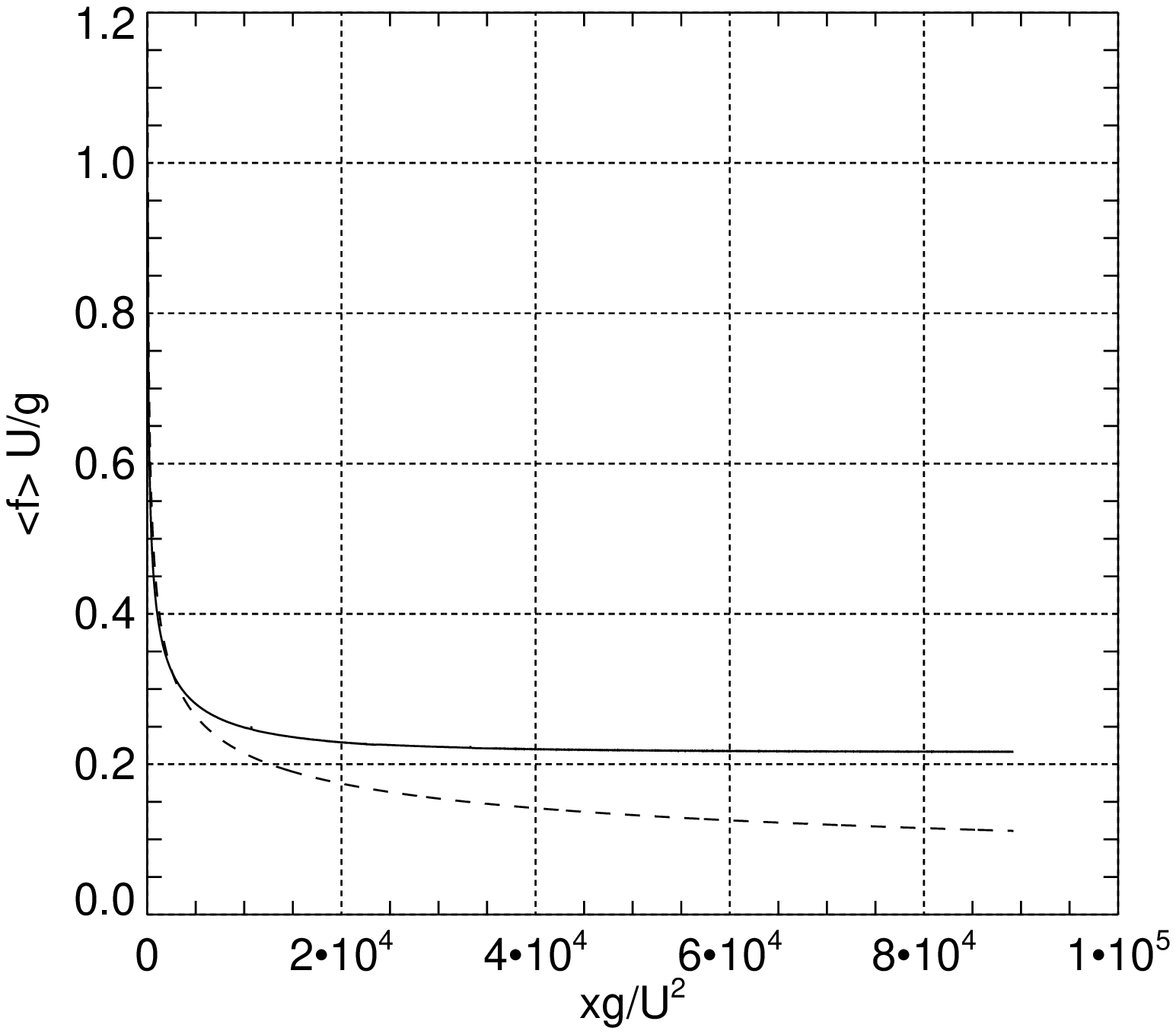}
                \caption{}
                \label{MeanFreqWAM3A}
        \end{subfigure}
\qquad
%(or a blank line to force the sub-figure onto a new line)
        \begin{subfigure}[b]{0.45\textwidth}
                \includegraphics[width=\textwidth]{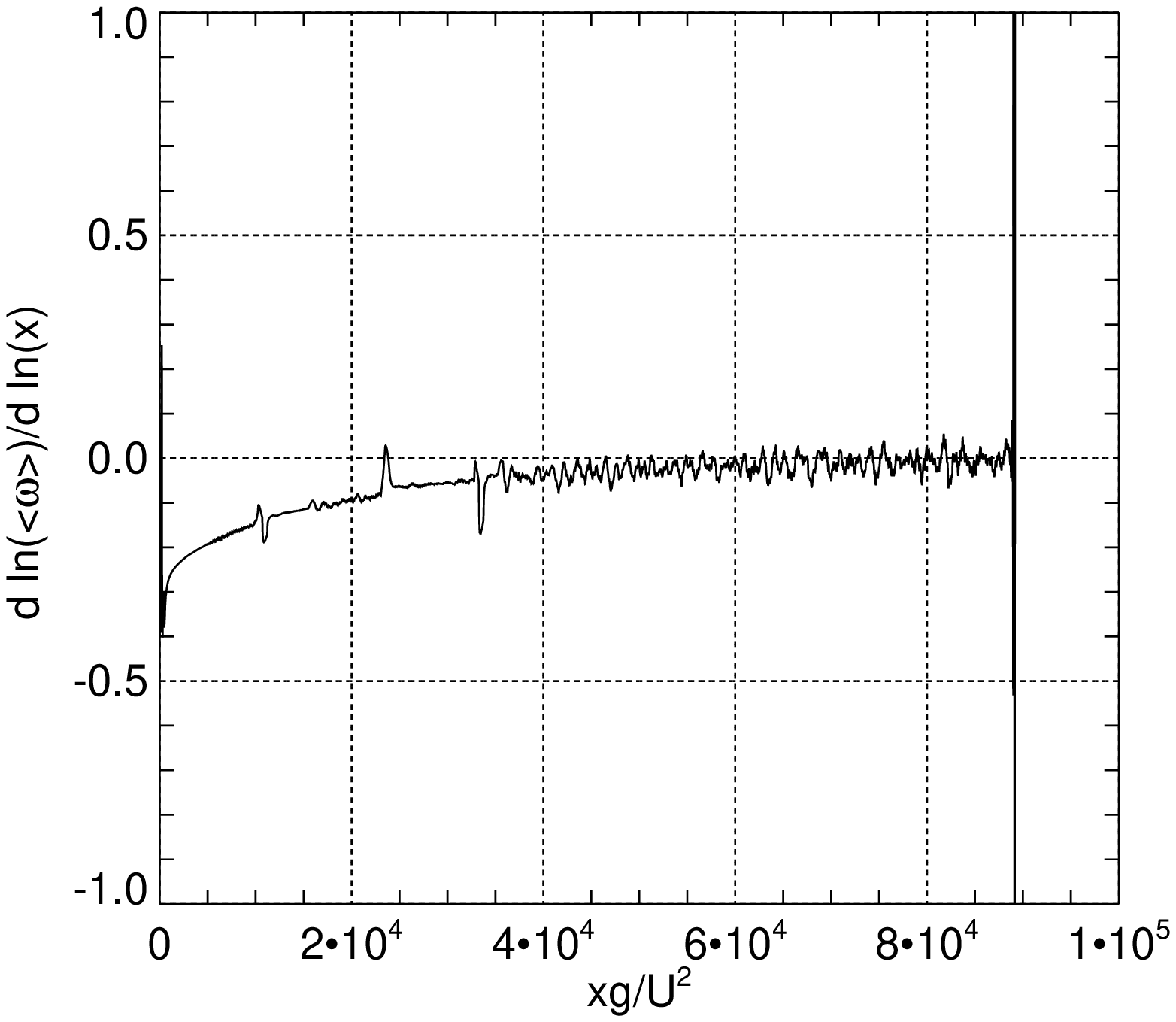}
                \caption{}
                \label{MeanFreqWAM3B}
	\end{subfigure}
	\caption{Same as Fig.\ref{MeanFreq}, but for $WAM3$ $S_{in}$}
	\label{MeanFreqWAM3}
\end{figure}

\noindent Similarly to energy, the dependence of the mean frequency, against the fetch, shown in Fig.\ref{MeanFreqWAM3A}, becomes constant at the dimensionless fetch value $\chi=5 \cdot 10^{4}$. The value of corresponding index $q$ goes asymptotically to $0$, see Fig.\ref{MeanFreqWAM3B}. Indicating discrepancies amongst $ZRP$ results.

%%%%%%%%%%%%%%%%%%%%%%%%%%%%%%%%%%%%%%%%%%%%%%%%%%%%%%%%%%%%%%%%%%%%%%%%%%%%%%%%%%%%%%%%%%%%%%%%%%

%%%%%%%%%%%%%%%%%%%%%%%%%%%%%%%%%%%%%%%%%%%%%%%%%%%%%%%%%%%%%%%%%%%%%%%%%%%%%%%%%%%%%%%%%%%%%%%%%%

\begin{figure}
        \centering
	\includegraphics[width=0.5\textwidth]{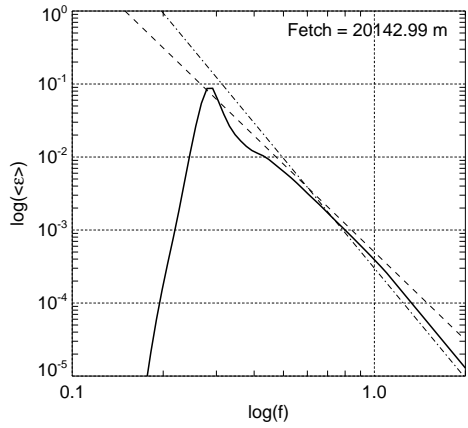}
	\caption{Same as Fig.\ref{SpecA}, but for $WAM3$ $S_{in}$.}
	\label{SpecWAM3}
\end{figure}

\noindent Nevertheless, Fig.\ref{SpecWAM3} demonstrates the Kolmogorov-Zakharov segment of the spectrum $\sim \omega^{-4}$, for small fetch value $\sim 20 \,\, km$.
%%%%%%%%%%%%%%%%%%%%%%%%%%%%%%%%%%%%%%%%%%%%%%%%%%%%%%%%%%%%%%%%%%%%%%%%%%%%%%%%%%%%%%%%%%%%%%%%%%

\begin{figure}
        \centering
	\includegraphics[width=0.5\textwidth]{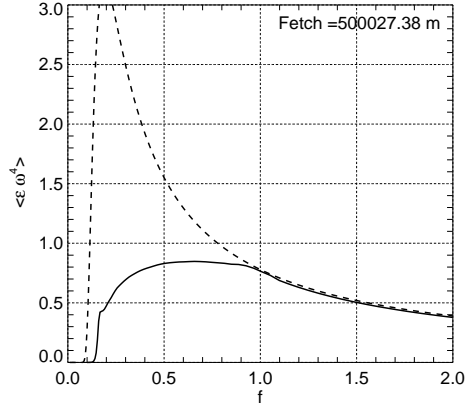}
	\caption{Angle averaged energy compensated spectrum $<\varepsilon \omega^{4}> = \frac{1}{2\pi}\int \varepsilon (\omega,\theta) \omega^{4} d \theta$ as the function of frequency $f=\frac{1}{2\pi} $  for $WAM3$ $S_{in}$.}
	\label{BetaKZ_WAM3}
\end{figure}

\noindent The solid line on Fig.\ref{BetaKZ_WAM3} presents, angle averaged, compensated wave energy spectrum, for stationary state, corresponding to dimensional fetch value $\sim 500 \,\, km$ (solid line). The dashed line demonstrates compensated Pierson-Moscowitz spectrum. Both spectra have coinsiding high-frequency behavior, but are completely different at lower $f$. Almost perfect correspondence at higher frequencies could mean that WAM3 input terms were tuned to match the experimental results only in spectral tail area.

%%%%%%%%%%%%%%%%%%%%%%%%%%%%%%%%%%%%%%%%%%%%%%%%%%%%%%%%%%%%%%%%%%%%%%%%%%%%%%%%%%%%%%%%%%%%%%%%%%

%%%%%%%%%%%%%%%%%%%%%%%%%%%%%%%%%%%%%%%%%%%%%%%%%%%%%%%%%%%%%%%%%%%%%%%%%%%%%%%%%%%%%%%%%%%%%%%%%%

\begin{figure}
        \centering
	\includegraphics[width=0.5\textwidth]{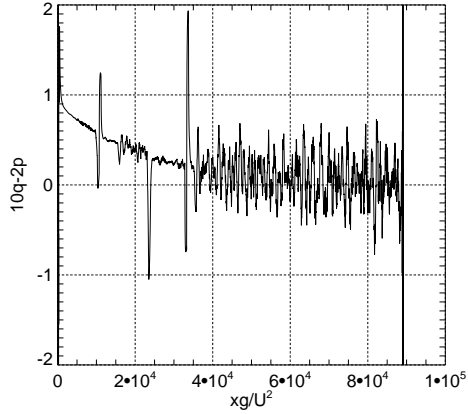}
	\caption{``Magic relation" $(10q-2p)$ as a function of the dimensionless fetch for $WAM3$ wind input term.}
	\label{MagicNumberWAM3}
\end{figure}

\noindent Fig.\ref{MagicNumberWAM3} presents combination $(10q-2p)$ as a function of the fetch. It is in total disagreement with the theoretical predictions. There is no indication of ``magic relation'' Eq.(\ref{MagicRelation}) fulfillment.

%%%%%%%%%%%%%%%%%%%%%%%%%%%%%%%%%%%%%%%%%%%%%%%%%%%%%%%%%%%%%%%%%%%%%%%%%%%%%%%%%%%%%%%%%%%%%%%%%%

%%%%%%%%%%%%%%%%%%%%%%%%%%%%%%%%%%%%%%%%%%%%%%%%%%%%%%%%%%%%%%%%%%%%%%%%%%%%%%%%%%%%%%%%%%%%%%%%%%
% not clear what 'it' is referring to, on the line below.
\noindent Comparison of $WAM3$ model with $JONSWAP$ experiments shows that it 
describes them fairly well, for small fetches of the order of $10 \,\, km$, i.e., in the region far from saturation and realization of the stationary state. Applicability of the model, for longer fetches, is questionable, at least for the reason that ``magic relation" ceases to be realized.

\section{Conclusion} \label{Conclusion}

\noindent Series of numerical experiments have been performed, for four different variants of wind input terms $S_{in}$, in the frame of the alternative numerical framework, which assumed:
\begin{enumerate}
\item exact nonlinear term $S_{nl}$
\item absence of spectral peak dissipation
\item ``implicit" high-wavenumbers dissipation in the form of Phillips tail  $\omega^{-5}$
\end{enumerate}

\noindent The fifth numerical experiment contained $WAM3$ spectral peak dissipation, ``by definition", but all other aspects of the numerical simulation correspond to the above alternative framework.

\noindent The results of all five numerical experiments were subjected to the five tests described below, with the summary presented in Table \ref{Table2}.

\begin{table}
\centering
%\begin{tabular}{|p{2.8in}|p{0.2in}|p{0.2in}|} \hline 
\begin{tabular}{ | l | l | l | l| l | l|} \hline
\textbf{Experiment} & $p$-test & $q$-test & $KZ$-spectrum & Magic relation & Energy growth \\ \hline 
$ZRP$ & YES & YES & YES & YES & YES \\ \hline 
$Chalikov$ & NO & NO & YES & YES & NO \\ \hline
$Snyder$ & NO & NO & YES & YES & NO \\ \hline 
$Hsiao-Shemdin$ & NO & NO & YES & YES & NO \\ \hline
$WAM3$ & NO & NO & YES & NO & NO \\ \hline
\end{tabular}
\caption{Summary of the tests performed on five models of wind input $S_{in}$}
\label{Table2}
\end{table}

\noindent The $p-$ and $q-$ tests are the checks for Eq.(\ref{Peq}) and Eq.(\ref{Qeq}) respectively; they check if the energy and mean frequency are power functions of the fetch, with proper self-similar exponents $p=1$ and $q=0.3$.

\noindent The $KZ$- spectrum test is a direct check of $WTT$'s validity, according to which, the directional (angle averaged) energy spectrum $<\varepsilon \omega^4>$ has to be, with up to $20\%$ accuracy, constant in the inertial interval $1.5\omega_p<\omega<3.5\omega_p$. Fulfilling this test directly points to the fact that $HE$, in the universal domain, is described by stationary Eq.(\ref{SnlZero}), which is caused by mutual cancellation of the ``in" and ``out" terms, as the dominating process in $S_{nl}$. 

\noindent The ``magic relation" test is a check for the ``magic relation" Eq.(\ref{MagicRelation}) and is more liberal than the $p-$ and $q-$ tests, since it assumes that power dependencies of the energy and mean frequency, along the fetch, are local, i.e. exponents $p = p(\chi)$ and $q=q(\chi)$ are slow functions of the fetch, but the ``magic relation" can still be fulfilled, for any value of the fetch coordinate.

\noindent The ``energy growth" test is a check that the energy growth rate, versus the fetch, compares with corresponding $ZRP$ dependence and the results of the 12 field experiments.

\noindent The following is a discussion of the above tests, applied to five simulations:

\begin{enumerate}

\item  It is no surprise that the $ZRP$ wind input function passed $p-$ and $q-$ tests, since it was especially designed with the purpose to satisfy them at $p=1.$ and $q=0.3$. It also passes $KZ$, ``magic relation'', and energy growth tests, since it reproduces more than a dozen of the field experiments. Therefore, it can serve as the benchmark.

\item  All other wind input terms also pass the $KZ-$ test. Validity of that result, for all five versions of the wind input, suggests its universality. One can say that, if not for all, then for a very wide choice of the wind input functions, the spectrum $\varepsilon \sim \omega^{-4}$ will be realized, due to domination of the conservative terms in the $HE$.

\item All the cases, except $WAM3$, passed the ``magic relation'' test. This means, practically, that for any form of quasi-linear $S_{in}$, for large fetches, there is formation of a local, self-similar, regime, with indices $p$ and $q$ slowly changing, with the fetch. It also confirms $WTT$.

\item  Chalikov and Hsiao-Shemdin cases fail the $p-$ and $q-$ tests, but are in qualitative agreement with the field experiments.

\item  All the cases, except $ZRP$, fail the energy growth test. 

\end{enumerate}

\noindent The Table \ref{Table1}, of section \ref{CheckingFramework}, presented the results of 12 experiments, confirming the law of $p=1$ and $q=0.3$ indices. Publication \cite{R6} presents the data of 24 field experiments - practically everything found in the literature. As it was already mentioned, half of those experiments satisfied $p-$ and $q-$tests. Simple calculation shows that ``magic relation" test is satisfied, with the accuracy of $30\%$, for $2/3$ of the described experiments. The reason for poor performance on the ``magic test", by other experiments, is discussed, in detail, in publication \cite{R6} and is explained, first of all, by data processing imperfections. More recent experiments, confirming the ``magic relation", are presented in publication \cite{R8}. One can conclude that the ``magic relation" is confirmed, nowadays, experimentally.

\noindent As far as the $WAM3$ model is concerned, while it does not pass any of the tests, except for $KZ$ spectrum test, it realizes speculative phenomenon of the ``mature sea", not confirmed by serious experiments, but described in publication \cite{R54}. In any case, none of the field experiments analyzed in \cite{R6}, \cite{R8} resemble, even remotely, formation of the ``mature sea". It would seem rational to refrain from this hypothesis.

\noindent The main obstacle for self-consistent wind-wave theory creation is the ambiguity of the analytical expression for $S_{diss}$. Although, it can be resolved, through a numerical solution of, not $HE$, but primordial Euler equations. Such experiments are already carried out and their result are partially published \cite{R56}, \cite{R57}. They unequivocally show that long wave dissipation occurs due to wave breaking of the short waves. Long wave dissipation, due to this process, is realized in rogue waves, but they are rare phenomena, and their contribution to energy balance is, at least, orders of magnitude lower than assumed in $WAM3$ and $WAM4$ models. This fact also justifies the author's lack of desire to supply the numerical model, by the long-wave dissipation.

\noindent Finally, let's discuss research perspectives. So far, the authors solved the $HE$, either as time evolution at a single spatial point \cite{R9}, \cite{R4}, \cite{R5} (``duration limited setup"), or as a spatially stationary solution \cite{R9} (``fetch limited setup"). Recent progress in the algorithms, computer software, and hardware development allow one to numerically solve the $HE$, using an exact $S_{nl}$ expression, for a fairly large amount of points in the temporal-spatial domain. Preliminary results on the $HE$ solution, for $40$ equidistant points, along the fetch, were already presented at the $WISE$ meeting \cite{R77}. Obtained results, for many more points (orders of magnitude larger) will be published soon.

\noindent It is expected that the plans to numerically simulate the wind wave development, using exact $S_{nl}$, are realistic, for fairly small domains like the Black Sea, Lake Michigan, and the Gulf of Bothnia.

\noindent The authors believe that the presented research is a step toward radical improvement of the existing operating models. The source functions $S_{in}$ and $S_{diss}$ must be carefully revised and optimized, by numerical experiments using the $XNL$'s nonlinear term. As soon as this is done, one should choose a proper approximation, for $XNL$ modeling -- some version of $DIA$, with optimally chosen quadruplets. Ideally, during forecasting, the model will not require parameter tuning.

\noindent The oceanographic society must start using solid, justified, physical models and abandon the ``black box  with tuning knobs" approach. A new model will require minimal, if any, tuning for different ocean conditions.

\section{Acknowledgments}

\noindent The authors would like to thank the referees of the manuscript for valuable discussion, which helped to improve the presented research.

\noindent The research presented in sections \ref{CurrentState} and \ref{TwoScenarios} has been accomplished due to support of the grant ``Wave turbulence: the theory, mathematical modeling and experiment"  of the Russian Scientific Foundation No 14-22-00174. The research presented in other chapters was supported by ONR grant N00014-10-1-0991. The authors gratefully acknowledge the support of these foundations.

\section{Appendix}

\noindent It is convenient to present the Hasselmann equation, not for the energy spectrum $\varepsilon(\omega,\theta)$, but for the wave action spectrum $n(k)$. The spectra are connected by the relation 

\begin{equation}
\omega_k n_k dk = \varepsilon(\omega,\theta) \omega d\theta
\end{equation}

The Hasselmann equation reads 

\begin{eqnarray}
\frac{\partial n_k}{\partial t} &=& 2 \pi g^2 \int_{k_2,k_3,k_4} (T_{k k_2 k_3 k_4})^2  \times   \\
&\times& (n_k n_{k_3} n_{k_4} + n_{k_2} n_{k_3} n_{k_4} -n_k n_{k_2} n_{k_3} -n_k n_{k2} n_{k_4}) \times  \nonumber \\
&\times& \delta(\omega+ \omega_2-\omega_3-\omega_4) \delta(k_1+k_2-k_3-k_4) dk_1 dk_2 dk_3 dk_4  \nonumber
\end{eqnarray}

\noindent Here $\omega_k = \sqrt{gk}$ and $T_{k_1 k_2 k_3 k_4} = \frac{1}{2} \left( \tilde{T}_{k_1 k_2 k_3 k_4} + \tilde{T}_{k_2 k_1 k_3 k_4}\right)$

\begin{eqnarray}
\label{MatrixElement}
T_{k_1 k_2 k_3 k_4} &=& \frac{1}{4} \frac{1}{(k_1 k_2 k_3 k_4)^{1/4}} \Bigg\{ \\
&+& \frac{1}{2} (k_{1+2}^2-(\omega_1+\omega_2)^4) \left[(\vec{k_1}\vec{k_2}-k_1 k_2) + (\vec{k_3}\vec{k_4}-k_3 k_4 \right] \nonumber \\
&-& \frac{1}{2} (k_{1-3}^2-(\omega_1-\omega_3)^4) \left[(\vec{k_1}\vec{k_3} -k_1 k_3) + (\vec{k_2}\vec{k_4}+k_2 k_4 \right] \nonumber \\
&-& \frac{1}{2} (k_{1-4}^2-(\omega_1-\omega_4)^4) \left[(\vec{k_1}\vec{k_4}+k_1 k_4) + (\vec{k_2}\vec{k_3}+k_2 k_3 \right] \nonumber \\
&+& \left( \frac{4(\omega_1+\omega_2)^2}{k_{1+2}-(\omega_1+\omega_2)^2}-1\right) (\vec{k_1}\vec{k_2}-k_1 k_2) (\vec{k_3}\vec{k_4}+k_3 k_4) \nonumber \\
&+& \left( \frac{4(\omega_1-\omega_3)^2}{k_{1-3}-(\omega_1-\omega_3)^2}-1\right) (\vec{k_1}\vec{k_3}+k_1 k_3) (\vec{k_2}\vec{k_4}+k_2 k_4) \nonumber \\
&+& \left( \frac{4(\omega_1-\omega_4)^2}{k_{1-4}-(\omega_1-\omega_4)^2}-1\right) (\vec{k_1}\vec{k_4}+k_1 k_4) (\vec{k_2}\vec{k_3}+k_2 k_3) \Bigg\} \nonumber
\end{eqnarray}
\noindent where notation $k_{1+2} = |\vec{k_1}+\vec{k_2}|$.

\noindent Eq.(\ref{MatrixElement}) can be rewritten as follows:

\begin{equation}
\frac{\partial n_k}{\partial t} = S_{nl} = F_k - \Gamma_k n_k
\end{equation}

\noindent where

\begin{eqnarray}
F_k &=& 2 \pi g^2 \int \left| T_{k k_1 k_2 k_3} \right|^2 n_{k_1}  n_{k_2} n_{k_3} \times \\
&\times& \delta(\vec{k}+\vec{k_1}-\vec{k_2}-\vec{k_3}) \delta(\omega_k+\omega_{k_1}-\omega_{k_2}-\omega_{k_3}) d k_1 d k_2 d k_3 \nonumber \\
\Gamma_k &=& 2 \pi g^2 \int  \left| T_{k k_1 k_2 k_3} \right|^2 (n_{k_1}  n_{k_2} +  n_{k_1}  n_{k_3} - n_{k_2}  n_{k_3}) \times \\
&\times& \delta(\vec{k}+\vec{k_1}-\vec{k_2}-\vec{k_3}) \delta(\omega_k+\omega_{k_1}-\omega_{k_2}-\omega_{k_3}) d k_1 d k_2 d k_3 \nonumber
\end{eqnarray}
\noindent $KZ$ stationary spectra are given by the expression

\begin{equation}
n_k = \frac{F_k}{\Gamma_k}
\end{equation}

\noindent As far as $n_k > 0$, $\Gamma_k > 0$. The stationary kinetic equation, in the presence of wind input and damping, reads

\begin{equation} 
S_{nl} + \gamma_k n_k = 0
\end{equation}

\noindent Here $\gamma_k = \gamma_{in} - \gamma_{diss}$. The solution to this equation is

\begin{equation}
n_k = \frac{F_k}{\Gamma_k - \gamma_k}
\end{equation}

\begin{figure}
	\center
	\includegraphics[width=0.5\textwidth]{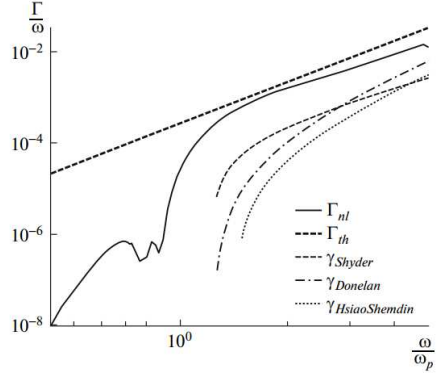}
	\caption{Coefficient of nonlinear dissipation $\Gamma_k$: theoretical estimate(dashed line) \cite{R1,R2} and numerical calculation (solid line). Known parameterizations, of the wave energy growth rate, are shown on the legend. Essential exceeding of $\Gamma_k$ over $\gamma_k$ explains applicability of the equation $S_{nl}$ = 0}. 
\label{Adopted}
\end{figure}

\noindent As far as $n_k > 0$, then $\Gamma_k > \gamma_k$. In fact, in the real situation $\Gamma_k >> \gamma_k$. This is clear from Fig.\ref{Adopted}, adopted from paper \cite{R2}. 

\noindent

\noindent It should be stressed out that our modernized code made possible separate calculation of $F_k$ and $\Gamma_k$.

\noindent Here the dashed line is the theoretical calculation of $\Gamma_k$, made narrow in the angle spectrum, and the solid line is the numerical experiment. More details can be found in paper \cite{R1}. 

\noindent Excess of $\Gamma_k$ over $\gamma_k$, by orders of magnitude, explains the dominant role of $S_{nl}$.

\section*{References}

\bibliography{mybibfile}

\end{document}